\documentclass[12pt,english]{article}
\usepackage[latin9]{inputenc}
\usepackage{geometry}
\usepackage{textcomp}
\usepackage{amsmath}
\usepackage{amssymb}
\usepackage{esint}
\usepackage[colorlinks, linkcolor=red, anchorcolor=blue, citecolor=blue]{hyperref}

\makeatletter
\usepackage{babel}
\usepackage{graphicx}
\usepackage{color} 

\begin{document}

%\date{}
%\title{Holograms of pure state black holes\\
%}

%\maketitle

%\begin{document}
\baselineskip 14 pt \parskip 12 pt

\begin{titlepage} 

\begin{flushright}
{\small{}RI-HUJI xx/15\\LMU-ASC 33/15} 
%{\small{}}
\par\end{flushright}

\begin{center}
\vspace{2mm}

\par\end{center}

\begin{center}
\textbf{\Large{}Hologram of a pure state black hole}\textbf{ }
\par\end{center}
\begin{center} 

Shubho R. Roy${}^{1,2}$ and Debajyoti Sarkar${}^{3,4}$

\vspace{3mm}

${}^1${\small \sl Racah Inst. of Physics} \\ {\small \sl Hebrew University of Jerusalem, Jerusalem 91904 Israel} \\ \smallskip {\small \tt shubho.roy@mail.huji.ac.il}

\vspace{3mm}

${}^2$\textsl{Theory Division}\\
\textsl{ }\textsl{\small{}Saha Institute of Nuclear Physics, Calcutta
700064, West Bengal, India}

\vspace{3mm}

${}^3${\small \sl Arnold Sommerfeld Center} \\ {\small \sl Ludwig-Maximilians-University, Theresienstr. 37, 80333 Munchen, Germany} \\ \smallskip {\small \tt debajyoti.sarkar@physik.uni-muenchen.de} 

\vspace{3mm}

${}^4${\small \sl Max-Planck-Institut f\"ur Physik, F\"ohringer Ring 6, D-80805 Munich, Germany}

\end{center} \vskip 0.8 cm 

\begin{abstract}
In this paper we extend the HKLL holographic smearing function method to reconstruct (quasi)local AdS bulk scalar observables in the background of a large AdS black hole formed by null shell collapse (a ``pure state" black hole), from the dual CFT which is undergoing a sudden quench. In particular, we probe the near horizon and sub-horizon bulk locality. First we construct local bulk operators from the CFT in the leading semiclassical limit, $N\rightarrow\infty$. Then we look at effects due to the finiteness of $N$, where we propose a suitable coarse-graining prescription involving early and late time cut-offs to define semiclassical bulk observables which are approximately local; their departure from locality being non-perturbatively small in $N$. Our results have important implications on the black hole information problem.
\end{abstract}
\end{titlepage}
\pagebreak{}

\section{Introduction\protect \\
}

The AdS/CFT duality \cite{Maldacena:1997re,Gubser:1998bc,Witten:1998qj,Aharony:1999ti} is, in principle, a fully non-perturbative definition of quantum gravity in asymptotically anti deSitter spacetimes (aAdS$_{d+1}$)  in terms of a \emph{large} $N$ factorizable conformal field theory supported on its conformal boundary (CFT$_d$). This definition is manifestly \emph{holographic} \cite{'tHooft:1993gx,Thorn:1991fv,Susskind:1994vu} but manifestly background dependent as well. One of the litmus tests for any such candidate theory of quantum gravity is the successful resolution of the black hole information paradox \cite{Hawking:1976ra}. AdS/CFT duality, of course, by definition, takes care of this paradox since any process in quantum gravity in aAdS$_{d+1}$, including formation and evaporation of black holes, is postulated to be described by a unitary dynamics of a local quantum field theory (CFT$_d$). However, this is true only at the level of principle, we are still in the process of understanding how in practice, quantum corrections in a CFT
 bring about restoration/recovery of bulk information from an evaporating black hole. The issue here is two-fold, first the mapping between gravitational degrees of freedom and gauge theory degrees of freedom is highly non-trivial and non-local and second, all semiclassical processes in the gravity side map to highly quantum and strongly coupled field theory dynamics. The usual approach of investigators motivated in issues on the gravity side, has been to sidestep the problem of solving a strongly coupled field theory, and just assume that a solution to strongly coupled field exists in principle but the details of the solution to field theory are not important. This is justified since black holes display a great degree of \emph{universality} and hence their dynamics should not depend on the specific details type of field theory, but instead only on broad-based or universal features of \emph{any} large N CFT. Then the crucial task becomes to properly map the gravitational degrees of freedom to the field theory degrees of freedom, which again should be independent of the detailed property of the specific CFT model/lagrangian.\\

Since majority of the interesting questions and issues on the gravity side (bulk) such as interiors of horizon or singularities are posed in terms of \emph{local} probes, we are interested in identifying CFT degrees of freedom which describe approximate local bulk gravitational physics\footnote{The original GKPW map \cite{Gubser:1998bc,Witten:1998qj} is not suitable for describing normalizable bulk fluctations in terms of which (quatum) local bulk physics is posed.}. Such a program was initiated since the early days of AdS/CFT \cite{Banks:1998dd,Balasubramanian:1999ri,Balasubramanian:1998sn, Bena:1999jv} and was brought to the best possible shape in the approach of Hamilton, Kabat, Lifschytz, Lowe (HKLL) and their collaborators \cite{Hamilton:2005ju,Hamilton:2006az,Hamilton:2006fh,Hamilton:2007wj,Lowe:2009mq,Kabat:2011rz,Kabat:2012hp,Kabat:2012av,Kabat:2013wga}. This map, known as the HKLL \emph{smearing} function in trade, reconstructs local bulk operators as delocalized operators in the CFT. Its construction is governed entirely by the CFT state (e.g. part of the conformal symmetries respected by the state), the particular representation of the CFT operator, and crucially, the requirement of bulk micro-causality. A natural consequence of these requirements is that the resultant bulk operator automatically satisfies bulk equations of motion derived from some local bulk lagrangian. In this sense this construction can be best thought as an \emph{inverse} LSZ construction i.e. recovering a local bulk field theory from asymptotic data.\\

The aim of this work is to construct CFT representations for \emph{quasi}local
operators in a collapsing black hole (BH) geometry in asymptotically AdS
background, first in the leading semiclassical approximation, and
then incorporating the effects of finiteness of $N$\footnote{In a CFT the factorization parameter is the central charge, $c$ which scales with the number of color/species as, $c\sim N^2$. However, here we are denoting the factorization parameter by $N$, in a somewhat abuse of notation. }.
The first part of work involves extending to arbitrary
higher dimensions, the smearing function construction of the two dimensional
(AdS$_{2}$) null-shell collapse example \cite{Lowe:2008ra} treated
in the leading semiclassical approximation ($N\rightarrow\infty$).
Such a construction for the collapse case was suggested in an earlier
work \cite{Hamilton:2006fh}. In \cite{Hamilton:2006fh} the smearing
function for eternal black holes in AdS$_{3}$ i.e. BTZ was worked
out with smearing support on a single boundary by analytic continuation
of boundary time to complex values. That construction \cite{Hamilton:2006fh},
in itself is highly significant, as local observables \emph{both}
outside and inside the event horizon are represented as delocalized
(smeared) operators in a \emph{single} CFT Hilbert space supported
on the \emph{sole} asymptotic boundary. However, for eternal AdS
black holes in arbitrary dimensions, there exists an alternative prescription
by Maldacena \cite{Maldacena:2001kr} to reconstruct the bulk as an
entangled state in the product Hilbert space of \emph{two} copies
of CFT's supported on each asymptotic boundary, called the \emph{thermofield double}
construction (TF) \cite{Israel:1976ur}. Such a construction entails access
to both CFT's to reconstruct all regions, inside and outside horizons
of the eternal black hole geometry. However for black holes formed
via collapse of a shell of matter, the second asymptotic region never
existed, and hence there are no alternative prescriptions to describe
the region behind the horizon using the ``second'' CFT. Our construction,
closes this existing gap in the literature regarding reconstruction
of single exterior black holes from the boundary CFT. \\

Next, we look at effects of finite (but still very large) $N$. This
is the most important part of the paper. For the eternal Schwarzschild
AdS (SAdS) black hole, effects of the finiteness of $N$ were captured
in the smearing construction through the \emph{late time cut-off}
proposed in \cite{Hamilton:2007wj,Lowe:2009mq,Kabat:2014kfa}. These
late time cut-offs were designed to capture the late time behavior
of Green's functions in the dual CFT at finite $N$ \cite{Birmingham:2002ph,Barbon:2003aq,Barbon:2004ce,Barbon:2005jr,Kleban:2004rx,Solodukhin:2005qy}.
At finite $N$, Green's functions start to depart from their thermal
expectation values as the CFT begins to sense the discreteness of
its spectra and further able to distinguish or discern individual
microstates. As such, bulk local operators inserted deep inside the
bulk which are smeared operators in the CFT with smearing support extending
to very late times, become only approximately local at large but finite
$N$. Local bulk operators which do not have smearing support at late
times can of course stay local even at very large but finite $N$.
Kabat and Lifschytz \cite{Kabat:2014kfa}, building on earlier work \cite{Hamilton:2007wj,Lowe:2009mq}, advocated a sensible way to capture
this aspect of a putative local bulk operator by introducing a late
time cut-off in the smearing integral. Smearing CFT operators over
time intervals (durations) any larger than this late time, say $t_{max}$
would seriously affect its bulk behavior and it would not be interpreted
as a local operator propagating in a semiclassical bulk black hole
spacetime. However for the collapse case, one also has to include
\emph{early time cut-off} effects of the finiteness of $N$. The
dual process in the CFT is the \emph{eigenstate thermalization} phenomenon
hypothesized by Deutsch \cite{PhysRevA.43.2046} and Srednicki \cite{PhysRevE.50.888}.
One must wait till the black hole pure state ``thermalizes'' i.e.
the CFT is no longer able to identify its precise microstate. At finite $N$ this time is non-zero and hence one needs to incorporate
this effect by inserting a early time cut-off, $t_{min}$%
\footnote{Here, in the CFT, we have a so called ``quench'' process - a \emph{far}
off-equilibrium CFT rapidly relaxing via strong self-coupling. And the time we are concerned about is the \emph{decoherence} time, after which the CFT is unable to recall its initial pure state (which is the precise quantum gravity microstate) and is well approximated by a coarse-grained state in gravity described by a semiclassical geometry. This time is also different from the quasinormal time scale (over which the black hole loses hair) as quasinormal phenomena is covered well inside the domain of local semiclassical bulk physics. %
}. In executing such a modification, the guiding philosophy is same
as that of Kabat and Lifschytz \cite{Kabat:2014kfa} -\\

 \emph{At finite (but large) $N$, one can only define local bulk operators approximately. This approximate local description is obtained by discarding the part of the CFT correlators which are sensitive to the detailed microstate/ pure state which collapses}.\\
 
The plan of the paper is as follows. In section \ref{sec:ESAdS_rev} we start off by reviewing the local bulk construction in eternal SAdS backgrounds and for bulk points both inside and outside the horizon in terms of both left and right CFT operators. Then following \cite{Hamilton:2006fh}, we review how to modify this construction using operators with support on only one boundary. In section \ref{sec:altEAdS} we present an alternative way to construct these bulk operators for eternal SAdS backgrounds which we will later use for the collapsing scenario. Sections \ref{sec:VaidyaAdS} and \ref{sec:VaidyaAdS_finteN} present our main results. There we use the machinery described in the previous sections to write down bulk operators at various regions of the AdS-Vaidya geometry and their finite $N$ modifications. Towards the end, in section \ref{sec:discuss} we discuss alternative approaches to bulk reconstruction for pure state black holes and highlight their similarities and differences with our construction. We comment on the implications of our work on black hole information problem/firewall paradox. Finally,in section \ref{sec:Outlook} we mention some open issues for future work.

%%%%%%%%%%%%%%%%%%%%%%%%%%%%%%%%%%%%%%%%%%%%%%%%
\section{Review of eternal SAdS bulk reconstruction\protect \\
}\label{sec:ESAdS_rev}
%%%%%%%%%%%%%%%%%%%%%%%%%%%%%%%%%%%%%%%%%%%%%%%%

Here we expand on the construction in appendix A of \cite{Kabat:2014kfa}
and especially, for points inside the event horizon where some features
might have been implicitly understood or assumed by the authors but
which we explicitly detail herein. As was pointed out in the previous section,
a manifestly 
\emph{regular} smearing function with \emph{compact} support only exists when boundary
spatial coordinates are continued to purely imaginary values - a \emph{spatial Wick Rotation} .
If one strictly adheres to real values of the boundary spatial coordinates,
the smearing functions do not exist in general \cite{Hamilton:2006fh},
and one has to work with singular smearing distributions \cite{Morrison:2014jha}.
A compromising alternative is to work with the smearing function in
(boundary) momentum space instead of position space \cite{Papadodimas:2012aq}.\\

The AdS Schwarzschild (SAdS$_{d+1}$) geometry is given by the metric,
\begin{equation}
ds^{2}=-f(r)dt^{2}+\frac{dr^{2}}{f(r)}+r^{2}d\Omega_{d-1}^{2},\label{eq: SAdS_d+1 metric}
\end{equation}
where $f(r)$ is the ``blackening function'', 
\[
f(r)=1+\frac{r^{2}}{R^{2}}-\left(1+\frac{r_{0}^{2}}{R^{2}}\right)\frac{r_{0}^{d-2}}{r^{d-2}}.
\]
$r_{0}$ is the horizon radius%
\footnote{Note that here in the metric we have traded the black hole mass parameter
$M$ for the horizon radius, $r_{0}$. They are related as,
\[
\frac{16\pi GM}{(d-1)S_{d-1}}=\left(1+\frac{r_{0}^{2}}{R^{2}}\right)r_{0}^{d-2}
\]
with $S_{d-1}$ is the volume of the $(d-1)$-sphere.%
} and $R$ is the AdS radius. $d\Omega_{d-1}^{2}$ is the metric on
$S_{d-1}$:
\[
d\Omega_{d-1}^{2}=d\theta^{2}+\sin^{2}\theta d\Omega_{d-2}^{2},\quad0<\theta<\pi.
\]
 \\
\begin{figure}[htbp] 
\begin{center} 
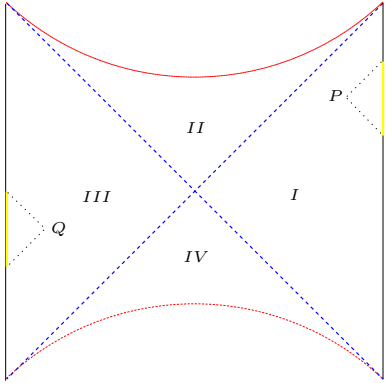 
\caption{The Penrose diagram of an eternal AdS Schwarzschild black hole. The red curved lines at the top and the bottom are the future and past curvature singularities. $P$ and $Q$ are local bulk operator insertions outside the event horizon and the yellow segments on the boundary are their respective boundary smearing function support.} 
\label{fig:SAdS_exterior} 
\end{center}
\end{figure}\\
The Penrose diagram for SAdS$_{d+1}$ is shown in figure ~\ref{fig:SAdS_exterior}.
Note that unlike the BTZ or AdS$_{2}$ black hole, the Penrose
diagram for SAdS in general dimensions is not a square; the past and
future singularities get bowed in \cite{Fidkowski:2003nf}.\\

It was shown in \cite{Kabat:2014kfa}, that a local scalar bulk operator,
$\Phi(r,t,\theta=0)$, inserted outside the horizon, i.e. $r>r_{0}$,
(indicated by the locations $P$ in quadrant $I$ and $Q$ in quadrant
$III$ in figure ~\ref{fig:SAdS_exterior}) can be reconstructed as
a boundary smearing of the dual CFT operator $\mathcal{O}$ (of appropriate conformal dimension $\Delta$), over
a region of the respective asymptotic boundary as follows:
\begin{equation}
\Phi(r,t,\theta=0,\Omega_{d-2})=\int_{\mbox{spacelike}}dt'd\phi'd\Omega'_{d-2}\: K(r,t,\theta=0,\Omega_{d-2}|t',\phi',\Omega'_{d-2})\:\mathcal{O}(t',\phi',\Omega'_{d-2}).\label{eq: Smearing for SAdS_d+1 outside horizon}
\end{equation}
Here, $\phi'\equiv i\theta$ is the imaginary angular coordinate, and the
domain of integration is of course over the boundary regions, spacelike
separated from the point of insertion of the bulk operator. Although
it appears strange at first to have boundary spatial coordinates such
as $\theta$ ``Wick rotated'' to purely imaginary coordinates $\phi'$,
it becomes very natural when we look what happens to the geometry
under this ``spatial'' Wick roation. Turning the angle $\theta$
to imaginary values as in $\phi$, converts the SAdS metric (\ref{eq: SAdS_d+1 metric})
into a metric which is asymptotically de Sitter%
\footnote{With an overall signature flip of the metric.%
}, and then the smearing function is given by the retarded Green's
function supported on the dS past infinity. Although an exact analytic
form of the SAdS smearing function is unknown, barring special cases
$d=1,2$, its existence is guaranteed by the fact that one can analytically
continue to an asymptotic de Sitter spacetime and that in de Sitter
an unique solution to the Cauchy problem can exist with past infinity
serving as the Cauchy surface. For our purposes the existence of the
smearing function is all we need, not the detailed analytic form or
a closed form expression in terms of special functions. In fact it
is known that the mode solutions to Klein-Gordon operator in SAdS background
cannot be solved analytically but only numerically (other than the
special cases, $d=1,2$)%
\footnote{For $d=4$, these modes are the Heun's functions, for which we do
not have any closed form expression either.%
}.\\

A noteworthy feature of this construction is that as the point of
insertion of the bulk operator is made to approach the future (past)
event horizon, the domain of the boundary smearing in the time direction
spreads all the way to the future (past) infinity. This is quite natural
as an event horizon is a global feature of a spacetime and for the
CFT to reconstruct it would require one to make CFT measurements to
all the way future or past, depending on whether it is a future horizon
or past horizon respectively. \\

%%%%%%%%%%%%%%%%%%%%%%%%%%%%%%%%%%%%%%%%%%%%%%%%%%%%%%%%%%%%%%%%%%
\subsection{Boundary representation of operator insertions within the horizon\protect \\
}
%%%%%%%%%%%%%%%%%%%%%%%%%%%%%%%%%%%%%%%%%%%%%%%%%%%%%%%%%%%%%%%%%%

\begin{figure}[htbp] 
\begin{center} 
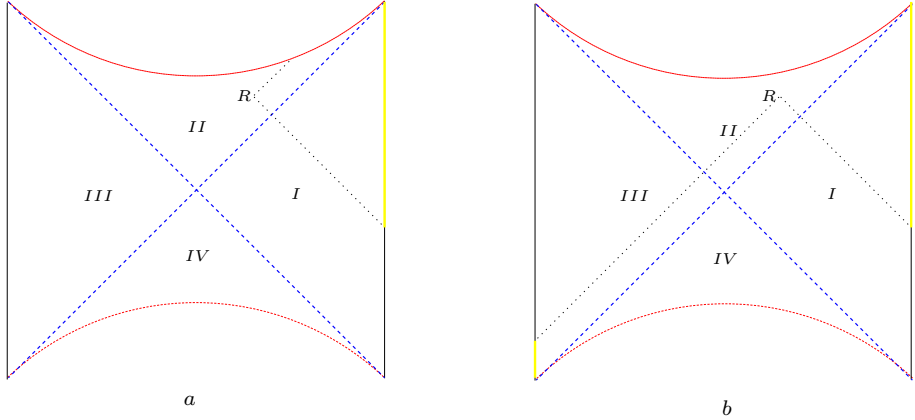 
\caption{The Penrose diagram of an eternal AdS Schwarzschild black hole with an operator inserted at $R$ inside the black hole horizon. a) Spacelike separated region to the right of the point $R$ exceeds the right boundary and spills over to the singularity. b) The smearing function support (yellow segments) of the inside horizon insertion, at $R$ - left boundary support is over timelike separated points and while right boundary support is over spacelike points. Similar smearing constructions are present for BTZ case as well as described below.} 
\label{fig:SAdS_interior} 
\end{center}
\end{figure}

For operator insertions inside the event horizon, say the point $R$
in quadrant $II$, indicated in figure ~\ref{fig:SAdS_interior}, the
above construction, which maps a local bulk operator to data on spacelike
separated points on the boundary, needs to be adjusted. This is because
the spacelike separated regions of $R$ on the right boundary, ``spills
over'' to a part of the curvature singularity, $r=0$ (see in figure ~\ref{fig:SAdS_interior} (a),
the region enclosed between the null lines emanating from $R$).\\
\\

For the spinless BTZ black hole or the AdS$_{2}$ ``black hole''
this situation was tackled in \cite{Hamilton:2005ju,Hamilton:2006fh,Hamilton:2007wj}
by using the ``antipodal map'' of the (global) AdS hyperboloid as
these black holes are quotients of pure AdS. One starts with Rindler
like coordinate patches of the global AdS. As in figure ~\ref{fig:SAdS_interior}, the global spacelike smearing
function support ``spills over'' to 
a ``post-singularity'' region in quadrant II of the BTZ-like/Rindler coordinate patch, but this spill over portion can be mapped back to the second asymptotic boundary using the antipodal
map of the Global AdS hyperboloid. There is no problem in using this
global AdS$_{3}$ smearing expression for BTZ - while taking the
quotient to form a black hole makes one periodically identify the spatial
boundary coordinate along the real axis, the smearing function
is only over purely \emph{imaginary} values of the boundary coordinates.
Suppressing the coordinates of the point $R$, then we have the following
boundary representation of the local operator placed at $R$:\footnote{The $z$ coordinate here (and also later on) is just a radial coordinate which puts the conformal boundary at $z\to 0$; not to be confused with the usual notation for Poincar\'{e} coordinate's radial direction.} 
\begin{equation}
\Phi_{BTZ}(R)=\int_{spacelike}dy\:\left[z\:\sigma(R|y,z)\right]^{\Delta-2}\mathcal{O}_{\Delta}^{R}(y)+\int_{timelike}dy'\:\left[-z'\:\sigma(R|y',z')\right]^{\Delta-2}(-)^{\Delta}\:\mathcal{O}_{\Delta}^{L}(y').\label{eq: BTZ inside horizon insertion smearning}
\end{equation}
Here $y(y')$ is condensed notation for all the right (left) boundary coordinates. The first integral is a smeared operator on the right boundary (the
superscript $R$ denotes that) and the smearing is over points on
the complexified right boundary which is spacelike separated. On the other hand, the second
integral is a smeared operator on the left boundary. This is shown
in figure ~\ref{fig:SAdS_interior} (b). Also $\sigma(R|y,z)$ is the AdS-invariant
``chordal distance'' while what we have in the expression, is the regulated
chordal distance which remains finite as the point $(y,z)$ is taken
to the boundary, $z\rightarrow0$. \\

The Global AdS$_{3}$ vacuum is of course non-invariant under the
quotienting as field modes must be periodic under the quotienting.
So instead of the AdS vacuum, we get the Kruskal or Hartle-Hawking
state, i.e. the state which is left invariant under the action of
the residual global symmetry%
\footnote{Quotienting makes the fields periodic along the quotient directions;
breaks down the global symmetry from $SO(2,2)$ to $SO(2)\times\mathbb{R}$.%
}.\\

However, higher dimensional eternal SAdS black holes are not quotients
of pure AdS and hence one cannot use the antipodal map%
\footnote{Although this trick of using antipodal maps would work for higher
dimensional hyperbolic black holes in AdS. We do not consider the
hyperbolic AdS black hole case as they are not suitable for collapse
situations unlike the SAdS case.%
}. Despite that, we expect a similar construction to also hold for
a higher dimensional SAdS black hole. A direct evidence for that
is the prescription of Maldacena \cite{Maldacena:2001kr} for the
eternal SAdS black hole in general dimensions. According to Maldacena,
the quantum gravity state of the eternal SAdS black hole aka the Hartle-Hawking
vacuum is given by ``thermofield double'' state which is a very
special entangled state of two identical CFT's supported on the two
asymptotic boundaries of the eternal SAdS geometry:
\begin{equation}
|\Psi\rangle_{Hartle-Hawking}=\sum_{n}e^{-\beta E_{n}/2}\:|E_{n}\rangle^{L}\otimes|E_{n}\rangle^{R}.\label{eq: thermofield double state}
\end{equation}
As such all correlators of supergravity operators in the Hartle-Hawking
vacuum must be reproduced by the correlators of their respective CFT
operator representations in the CFT entangled state. Indeed, it was
shown in \cite{Hamilton:2006fh,Hamilton:2007wj} that for $d=2$,
the CFT smeared representation (\ref{eq: BTZ inside horizon insertion smearning}),
reproduces supergravity correlators in the Hartle-Hawking vacuum.
So we combine insights and propose an ansatz of the form parallel
to (\ref{eq: BTZ inside horizon insertion smearning}), for inside
horizon operators in the background of general dimensional SAdS black holes,
\begin{equation}
\Phi(R)=\int_{spacelike}dy\; K_{SAdS}(R|y)\:\mathcal{O}_{\Delta}^{R}(y)+\int_{timelike}dy'\: K_{SAdS}(R|y')f(\Delta,R,y')\:\mathcal{O}_{\Delta}^{L}(y').\label{eq: CFT representation of inside horizon operator insertion in Eternal SAdS}
\end{equation}
The choice of the support of the smearing to be restricted to timelike
separated points on the other/left boundary is motivated by the fact
that for an outside horizon bulk operator the support on the far boundary
must vanish entirely \footnote{However, this left-side smearing function with a timelike support should not be construed as being derived from a retarded Green's function through a Green's theorem. In fact, recall that in the BTZ case of \cite{Hamilton:2006fh,Hamilton:2007wj} this left-side smearing was derived from a spacelike smearing from the ``spillover" portion of the right boundary across the singularity using the anti-podal map.}. The function, $f(\Delta,R,y')$ is hitherto
unspecified but can be worked out by demanding that it be such
that the CFT representation (\ref{eq: CFT representation of inside horizon operator insertion in Eternal SAdS})
reproduce the supergravity correlators. For example, consider the
correlator of this ansatz operator with an operator on the right boundary.
For the smearing construction to be compatible with Maldacena \cite{Maldacena:2001kr},
\begin{eqnarray*}
\langle\Phi(R)\mathcal{O}_{\Delta}^{R}(w)\rangle_{Hartle-Hawking} & = & \int_{spacelike}dy\; K_{SAdS}(R|y)\:\langle\mathcal{O}_{\Delta}^{R}(w)\mathcal{O}_{\Delta}^{R}(y)\rangle_{TF}+\\
 &  & \int_{timelike}dy'\: K_{SAdS}(R|y')f(\Delta,s(R,y'))\:\langle\mathcal{O}_{\Delta}^{L}(w)\mathcal{O}_{\Delta}^{R}(y')\rangle_{TF},
\end{eqnarray*}
Here we have used the subscript $TF$ to denote correlators in the
thermal entangled state of the two CFT's while $s(R,y')$ is the \emph{regulated}
geodesic distance from the point of insertion $R$ to a point on
the complexified left boundary with coordinate $y'$%
\footnote{The geodesic distance between a point in the bulk, say $R$ and a
point with coordinates$(z',y')$ approaching the asymptotic AdS boundary
diverges as some powers of radial coordinate, in this case $z'$. Just as in
the case for $BTZ$, we need to multiply the geodesic length by appropriate
factors of the radial coordinate $z'$ of the point which is approaching
the boundary, to define a regulated geodesic distance. For pure AdS,
this is well-known, $s\sim\lim_{z'\rightarrow 0}z'\sigma(z,X|z',X')$
 where $\sigma$ is the AdS-invariant chordal distance.%
}. One can in principle determine the left hand side of the above equation from
the supergravity computations by taking the (right) boundary limit
of the second operator which is an outside horizon insertion, 
\[
\langle\Phi(R)\mathcal{O}_{\Delta}^{R}(w)\rangle_{Hartle-Hawking}=\lim_{z\rightarrow0}\: z^{-\Delta}\,\langle\Phi(R)\Phi(z,w)\rangle_{Hartle-Hawking},
\]
while regarding the right hand side (rhs), one has knowledge of the thermofield correlators
from thermal CFT computations and $K_{SAdS}$ (from continuation from
de Sitter), as well as the domain of integrations. Enforcing this equality
lets one figure out the hitherto $f(\Delta,s(R,y'))$, in principle
(For BTZ, $f=f(\Delta)=(-)^{2\Delta-2}$, i.e. a pure phase). We will
not try to compute $f$ explicitly as the integrals on the rhs cannot
be done analytically. As it has been mentioned before, the analytic expressions
for $K_{SAdS}$ are unavailable, and we will just be content with the
knowledge that such a solution exists in principle.\\

%%%%%%%%%%%%%%%%%%%%%%%%%%%%%%%%%%%%%%%%%%%
\subsection{Eternal black holes in a single CFT\protect \\
}
%%%%%%%%%%%%%%%%%%%%%%%%%%%%%%%%%%%%%%%%%%%

Following Maldacena \cite{Maldacena:2001kr}, it has become customary
to view that the eternal AdS black hole is the dual to the thermally
entangled pure state of two identical CFT's called the \emph{thermofield double}
state. These two CFT's are supported on the two spatial infinities (boundaries)
of the two asymptotic regions of the eternal SAdS geometry. However
in our view it is better to think of the eternal SAdS black hole as the
dual geometry corresponding to a \emph{single} CFT, but in a mixed
(thermal) state - described by the canonical ensemble density operator.
This might seem to be a very superficial difference in effect, but
in concept there is a significant shift. An elementary result in quantum
mechanics is that every mixed state or mixed density matrix can have
several ``purifications''. This means one can express the mixed
state density matrix of the system as a reduced density matrix when
one considers the original system, say $A$ to be a part of a larger
Hilbert space, to wit, $A\times B$ (where $B$ is an auxiliary Hilbert
space) and subsequently ``tracing over'' $B$. Maldacena's picture
of the eternal SAdS black hole then becomes one of the many possible
purifications of a thermal state but this particular purification
is especially appealing because the auxiliary Hilbert space, $B$ has
a natural support on the spatial boundary of the second asymptotic
region. However this logic would become cumbersome if we were to apply
it to a charged or spinning black hole which usually has infinite numbers of asymptotic regions. This picture would compel us to describe the charged
or spinning black hole as a dual to entangled states over infinite
copies of CFT's, supported on the boundaries of these infinite asymptotic
regions. So instead, we insist on viewing the eternal SAdS black hole
as a thermal mixed state of a single CFT. The fact that putting the
CFT at finite temperature does not change the spectrum means we can
use the same operators as in the zero temperature version, to create
and destroy quanta. Then it becomes obvious that excitations of such
a system at finite temperature, such as local operators \emph{everywhere}
in the eternal SAdS geometry, including those behind the horizons,
can be accomplished by operators acting on states of single CFT/ boundary.
Maldacena's purification of thermal state of single CFT in terms of
a pure entangled state of two separate CFT's have somehow led to
the general impression that excitations in the eternal SAdS geometry,
in particular sub-horizon excitations cannot be described in a single
CFT. However to highlight the conceptual parallel to the collapse
black hole situation, we emphasize the point of describing the entire
eternal SAdS geometry as thermal state of a single CFT, because the
CFT pure state (which forms the black hole) would look like a thermal
state in a short time (eigenstate thermalization time) i.e. like an eternal
AdS geometry. \\

\begin{figure}[htbp] 
\begin{center} 
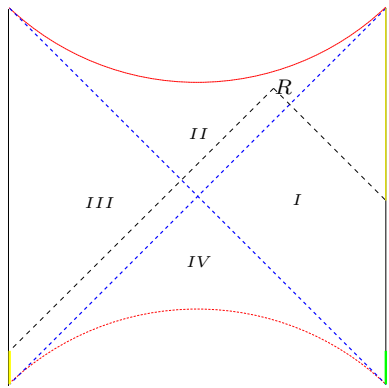 
\caption{The smearing function support for sub-horizon operators. Using complexified Schwarzschild time one can bring over the smearing integral over the left boundary, shown in yellow at the bottom left corner, to its symmetric region on the right boundary, shown in green at the bottom right corner.} 
\label{fig:SAdS_left_right} 
\end{center}
\end{figure}

To develop things further, we re-express the smearing expressions
for sub-horizon operator insertions in (\ref{eq: CFT representation of inside horizon operator insertion in Eternal SAdS})
on a single boundary. The way to accomplish this is by promoting the
Schwarzschild time coordinate from a local real coordinate defined
only in region $I$ to a global but \emph{complex} coordinate following
\cite{Fidkowski:2003nf}. On crossing horizon the global Schwarzschild
time coordinate picks up imaginary parts. The imaginary parts in the
four quadrants of any maximally extended Schwarzschild geometry can
be chosen as follows:
\begin{equation}
Im(t)=\left\{ \begin{array}{cc}
0, & I\\
-i\:\beta/4 & II\\
-i\:\beta/2 & III\\
+i\:\beta/4 & IV
\end{array}\right.\label{eq: Imaginary parts of Schwarzschild time in 4 quadrants of an eternal black hole.}
\end{equation}
Here $\beta$ is the inverse Hawking temperature. For the eternal
SAdS black hole, 
\begin{equation}
\beta=4\pi/\left(\frac{d\: r_{0}}{R^{2}}+\frac{d-2}{r_{0}}\right).\label{eq: Hawking temperature inverse of the SAdS black hole}
\end{equation}
The real part of the Schwarzschild time also reverses direction, when
one moves from quadrant $I$ to quadrant $III$. A result of adopting
such a complex global Schwarzschild time coordinate is that correlators of (normalizable) SUGRA operators
supported on quadrant $III$, $\Phi^{L}(t,\Omega_{d-2})$ in the Kruskal or Hartle-Hawking state
is identical to correlators of SUGRA operators on the right on quadrant $I$ with right time coordinate
shifted in the imaginary direction%
:
\begin{equation}
\langle \Phi^{L}(r,t,\Omega_{d-2})\dots\rangle_{Hartle-Hawking}=\langle \Phi^{R}(r,t-i\beta/2,\Omega_{d-2})\dots\rangle_{Hartle-Hawking}.\label{eq: Identification of SUGRA operators on left and right quadrants}
\end{equation}
Since AdS/CFT dictionary identifies boundary limits of normalizable operators with operators on the CFT, we can substitute the left boundary CFT operators acting on the thermofield state by the action of right boundary operators with analytically extended time coordinate\footnote{In a CFT at finite temperature, we study real-time thermal Wightman
functions 
\[
G_{RR}(t,\Omega_{d-2};0)=\mbox{Tr}\left[e^{-\beta H}\:\mathcal{O}(t,\Omega_{d-2})\:\mathcal{O}(0,0)\right]
\label{eq: left operators as right}\]
In our case this is the gauge theory supported on the right boundary,
which explains the subscript $RR$. If we analytically continue these
right boundary gauge theory Wightman functions to complex time, we
compute the correlation function of operators inserted on opposite
asymptotic boundaries 
\[
G_{LR}(t,\Omega_{d-2};0)=\mbox{Tr}\left[e^{\text{\textminus}\beta H}\:\mathcal{O}(t-i\beta/2,\Omega_{d-2})\:\mathcal{O}(0,0)\right]=G_{RR}(t-i\beta/2,\Omega_{d-2};0).
\]

It is crucially important to note that as we are working at the level of the Green's functions, these resulting bulk-boundary relations are to be understood inside a correlation function and not as an operator equations. An operator level identity is much harder to come by and progress in this direction has been made recently in \cite{Papadodimas:2015jra} and their follow-ups, where it has been argued that a notion of ``state dependence'' in quantum gravity is necessary to obtain a rigorous bulk-boundary operator correspondence. This is still a subject of ongoing research, which we won't be discussing here further; although see section \ref{sec:discuss} for a brief discussion.}
:
\begin{equation}
\langle \mathcal{O}^{L}(t,\Omega_{d-2})\ldots\rangle_{TF}=\langle\mathcal{O}^{R}(t-i\beta/2,\Omega_{d-2})\ldots\rangle_{TF}.\label{eq: Identification of operators on left and right boundaries}
\end{equation}
 Using this mapping, the boundary smearing representation of sub-horizon
operator insertion in (\ref{eq: CFT representation of inside horizon operator insertion in Eternal SAdS})
is re-expressed in terms of operators supported on the right boundary
continued to complex time:
\begin{equation}
\Phi(R)=\int_{spacelike}dy\; K_{SAdS}(R|y)\:\mathcal{O}_{\Delta}^{R}(y)+\int_{y'_{min}}^{\infty}dy'\: K_{SAdS}(R|y')f(\Delta,R,y')\:\mathcal{O}^{R}_{\Delta}(y'^{0}-i\beta/2,\Omega_{d-2})\label{eq: Single boundary CFT representation of inside horizon operator insertion in Eternal SAdS}
\end{equation}
Here $y'_{min}$ is the Schwarzschild time coordinate of the point
where a null ray from $R$ hits the left boundary. Changing variables,
$y'\rightarrow-y'$, we obtain,
\begin{equation}
\Phi(R)=\int_{spacelike}dy\; K_{SAdS}(R|y)\:\mathcal{O}^{R}_{\Delta}(y)+\int_{-\infty}^{-y'_{min}}dy'\: K_{SAdS}(R|-y')f(\Delta,R,-y')\:\mathcal{O}^{R}_{\Delta}(-y'^{0}-i\beta/2).\label{eq: Single CFT representation of inside horizon operator insertion 2 in Eternal SAdS}
\end{equation}
However the second integral may now be thought of as over the right
boundary with complexified time (and space). This is depicted in figure ~\ref{fig:SAdS_left_right}.
Note that since the right boundary smearing is over a time range with
a non-vanishing imaginary part, the figure is not entirely accurate.
We have just shown the real part of the time range on the right boundary
smearing (green). The imaginary time direction is perpendicular to
the plane of the paper.
\\

This expression could be put to use to represent fields in the single
exterior black holes, such as the $\mathbb{RP}^{2}$ geon \cite{Louko:1998hc}.
There is no physical second/left boundary in that case, however we
can always \emph{define} analogous operators within a single (right)
boundary CFT using the map (\ref{eq: Identification of operators on left and right boundaries}) because the state of the system allows us to do so. This is how or why state dependence is fundamentally built into the bulk-boundary map.\\

\begin{figure}[htbp] 
\begin{center} 
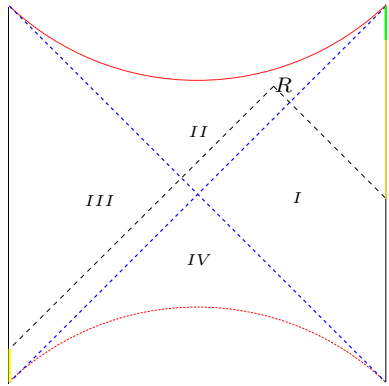 
\caption{The smearing function support for sub-horizon operators. The contribution of the region shown in green at the bottom i.e $t<0$ of figure ~\ref{fig:SAdS_left_right} can be time-evolved and converted into a integral in the $t>0$ half, shown in this figure in green again.} 
\label{fig:SAdS_left_right2} 
\end{center}
\end{figure}

For future reference, we note here that using boundary time evolution,
we have,
\[
O\left(-t-i\beta/2\right)=e^{i2Ht}O(t-i\beta/2)e^{-i2Ht}.
\]
Here, $H$ is the undeformed CFT Hamiltonian. Using this forward time evolution, we can bring the contribution
from the green segment in figure ~\ref{fig:SAdS_left_right} which is in the $t<0$ half to $t>0$ half. This is shown in figure ~\ref{fig:SAdS_left_right2}.

After this evolution, we can re-express a bulk sub-horizon bulk insertion in (\ref{eq: Single CFT representation of inside horizon operator insertion 2 in Eternal SAdS}) in terms of time-evolved operators accompanied by a flip of sign of time variable, $y'$:
\begin{equation}
\Phi(R)=\int_{spacelike}dy\; K_{SAdS}(R|y)\:\mathcal{O}^{R}_{\Delta}(y)+\int_{y'_{min}}^{\infty}dy'\: K_{SAdS}(R|y')f(\Delta,R,y')\:e^{i2Hy'^{0}}\mathcal{O}^{R}_{\Delta}(y'^{0}-i\beta/2)e^{-i2Hy'^{0}} \label{eq. No trans-Planckian}
\end{equation}

%%%%%%%%%%%%%%%%%%%%%%%%%%%%%%%%%%%%%%%%%%%%
\section{An alternative eternal SAdS smearing\protect \\
}\label{sec:altEAdS}
%%%%%%%%%%%%%%%%%%%%%%%%%%%%%%%%%%%%%%%%%%%%

\begin{figure}[htbp] 

\begin{center} 
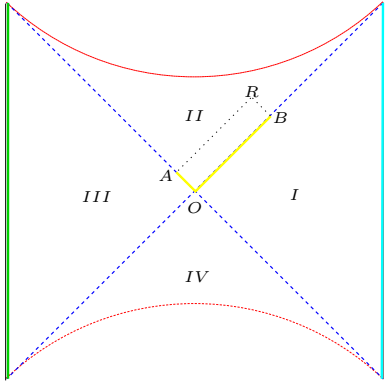 
\caption{The alternative smearing function support for sub-horizon operators.} 
\label{fig:SAdS_inside_2} 
\end{center}
\end{figure}

To prepare the ground for our treatment of the collapse situation, we present
an alternative smearing function for the eternal SAdS. This new construction
is identical to the Kabat-Lifschytz construction \cite{Kabat:2014kfa}
for local bulk operator insertions at points outside the horizon such
as quadrant $I$, but it differs for operator insertions in the interior
quadrant $II$ as indicated in figure ~\ref{fig:SAdS_inside_2}. \\
\\

A local bulk operator at $R$ is first expressed as an integral over
the horizons ($u=0\cup v=0$ in SAdS Kruskal coordinates \cite{Maldacena:1997re})
using retarded Green's functions for SAdS geometry.
\begin{eqnarray}
\Phi(R) & = & \int_{u\rightarrow0}\sqrt{h}dv\, d\Omega'_{d-1}\:\left[G_{SAdS}(R|u,v)\overleftrightarrow{\nabla}{}_{u}\Phi(u,v)\right]\nonumber \\
 &  & \qquad+\int_{v\rightarrow0}\sqrt{h}du\, d\Omega'_{d-1}\:\left[G_{SAdS}(R|u,v)\overleftrightarrow{\nabla}{}_{v}\Phi(u,v)\right].\label{eq: SAdS sub horizon intermediate}
\end{eqnarray}
\\
This is indicated in the figure by the yellow segments $OA$ (on $v=0$)
and $OB$ (on $u=0$). $h$ is the induced metric on the $u(v)=0$
surfaces, and $\Omega'$ is the usual solid angle on the $S_{d-1}$
on the horizons (suppressed in the figure). Then each integral over
the horizon is re-expressed as integrals over respective boundaries
(green and cyan) using the usual SAdS smearing, i.e. we plug the following
expressions 
\[
\lim_{u\rightarrow0}\Phi(u,v)=\int_{Right}dt\, d\Omega''_{d-1}\: K_{SAdS}(u,v|t,\Omega''_{d-1})\:\mathcal{O}^{R}(t,\Omega''_{d-1}),
\]
\[
\lim_{v\rightarrow0}\Phi(u,v)=\int_{Left}dt\, d\Omega''_{d-1}\: K_{SAdS}(u,v|t,\Omega''_{d-1})\:\mathcal{O}^{L}(t,\Omega''_{d-1}),
\]
in the previous expression (\ref{eq: SAdS sub horizon intermediate})
and obtain,
\begin{eqnarray}
\Phi(R) & = & \int_{Right}dt\, d\Omega''_{d-1}\:\left(\int_{u\rightarrow0}\sqrt{h}dv\, d\Omega'_{d-1}\: G_{SAdS}(R|u,v)\overleftrightarrow{\nabla}{}_{u}K_{SAdS}(u,v|t,\Omega''_{d-1})\right)\:\mathcal{O}^{R}(t,\Omega''_{d-1})\nonumber \\
 \qquad &  & +\int_{Left}dt\, d\Omega''_{d-1}\:\left(\int_{v\rightarrow0}\sqrt{h}du\, d\Omega'_{d-1}\: G_{SAdS}(R|u,v)\overleftrightarrow{\nabla}{}_{v}K_{SAdS}(u,v|t,\Omega''_{d-1})\right)\:\mathcal{O}^{L}(t,\Omega''_{d-1})\nonumber \\
\label{eq: Alternative SAdS sub-horizon}
\end{eqnarray}
Notice now the smearing regions are completely non-compact (both to
the past and future). When $R$ approaches either of the horizons,
say $u=0$, then it is clear from the figure that the intercept, $OA$
on the other horizon, $v=0$ vanishes and thus one has a smearing
representation on a single (right) boundary. Note that in such limits, i.e. when the point $R$ approaches $u=0$ ($v=0$), we are left with only the first term (second term) on the rhs of (\ref{eq: Alternative SAdS sub-horizon}). But note that in this limit, we also know how to represent the point $R$ as a smeared boundary operator over compactified right (left) bondary region. Thus we get an identity between the integrals appearing in (\ref{eq: Alternative SAdS sub-horizon}) and the usual smearing integrals. These identities can be used to justify our constructions (and to some extent can serve as a partial proof) in the next sections, where we apply this technique to AdS-Vaidya geometry.\\

%%%%%%%%%%%%%%%%%%%%%%%%%%%%%%%%%%%%%%%%%%%%%%%%%%%%
\section{Black holes by collapse: the Vaidya-AdS spacetime\protect \\
}\label{sec:VaidyaAdS}
%%%%%%%%%%%%%%%%%%%%%%%%%%%%%%%%%%%%%%%%%%%%%%%%%%%%

\begin{figure}[htbp] 
\begin{center} 
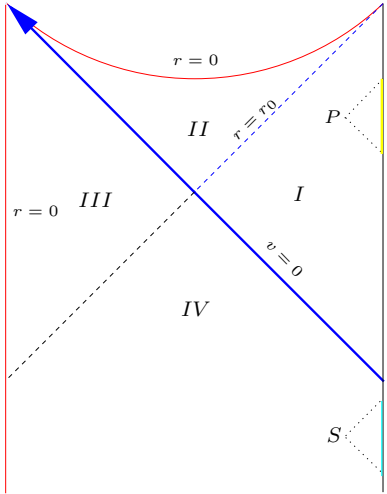 
\caption{The Penrose diagram of the Vaidya AdS spacetime. The trajectory of the imploding null shell,  is shown in blue (solid). The (future) singularity and the center of global AdS, both at $r=0$ are represented by red segments. The dashed blue segment is the (future) event horizon. $P$ and $S$ are local operator insertions.} 
\label{fig:Vaidya-AdS} 
\end{center}
\end{figure}

The eternal black hole, while being conceptually simpler, is too exotic
for realistic physical situations. Such a geometry cannot be produced
via gravitational collapse. In order to gain insight into how holography
informs the dynamics of quantum gravity one needs to consider more
physical situations where a black hole forms out of gravitational
collapse of matter-energy. One such simple example of a collapse geometry
is the Vaidya-AdS spacetime. The Vaidya-AdS spacetime is constituted
by taking the right ($I$) and future ($II$) regions of an eternal
SAdS$_{d+1}$ black hole and joining them across the in-falling null
shell to a piece of AdS$_{d+1}$. The Penrose diagram of the Vaidya-AdS
geometry is shown in figure ~\ref{fig:Vaidya-AdS}. \\
\\
The imploding shell is described by a spherically symmetric ``null
shock'' i.e. the Vaidya geometry is a solution to Einstein's equation
with the shell stress-tensor of the form of a Dirac delta function
supported on an in-falling null-ray, $v=0$. The metric is expressed
as:
\[
ds^{2}=-f(r,v)dv^{2}+2dvdr+r^{2}d\Omega_{d-1}^{2},
\]
where $d\Omega_{d-1}^{2}$ is the metric on the $S_{d-1}$. The function
$f(r,v)$ is discontinuous across the shock, $v=0$:
\[
f(r,v)=1+\frac{r^{2}}{R^{2}}-\theta(v)\left(1+\frac{r_{0}^{2}}{R^{2}}\right)\frac{r_{0}^{d-2}}{r^{d-2}}.
\]
Here $r_{0}$ is the horizon radius and $R$ is the AdS radius. We
are interested in forming a ``large'' black hole, for which $r_{0}>R$.\\

Note that the Vaidya-AdS spacetime is divided into four distinguishable
regions, indicated by the roman numerals, $I-IV$ in the Penrose diagram,
depending on whether they are to the past or future of the shell and
whether they lie to the interior or exterior to the horizon, $r=r_{0}$.%
\footnote{Of course for the region $III$ i.e. $v<0\cap r<r_{0}$, there is no
horizon as the shell has not yet collapsed inside its Schwarzschild
radius. Nevertheless, it is still a trapped region, and we will keep
referring to region $III$ as to the ``past of the shell, inside the
horizon'' without any confusion or inconsistency.%
}

For each coordinate patch, one can introduce a local tortoise coordinates,
$(r^{*},t)$ defined by,
\begin{eqnarray}
r^{*} & = & \int^{r}\frac{dr}{f(r)}+C\nonumber \\
t & = & v-r^{*}.\label{eq: tortoise coordinates}
\end{eqnarray}
 
We shall discriminate the local tortoise coordinates by use of different
subscripts $\left(r_{f}^{*},t_{f}\right)$ for $v>0$ and $\left(r_{i}^{*},t_{i}\right)$
for $v<0$. In the tortoise coordinates the Vaidya-AdS metric assumes
the familiar form,
\[
ds^{2}=-f(r(r^{*}))\left(dt^{2}+dr^{*2}\right)+r^{2}(r^{*})d\Omega_{d-1}^{2}.
\]

Note that in these coordinates lightcones are shifted across $v=0$
in a Penrose diagram i.e. the tortoise coordinates are discontinuous
across $v=0$ in general. However we shall choose the integration
constants, $C_{i/f}$ in defining $r^{*}$ so that they agree at the
asymptotic timelike infinity, 
\[
r_{f}^{*}=r_{i}^{*}=0\quad \mbox{at}\quad r\rightarrow\infty.
\]
This automatically ensures, $t$'s are also continuous across the
shock at asymptotic boundary. 

\[
t_{i}(v\rightarrow0^{-},r\rightarrow\infty)=t_{f}(v\rightarrow0^{+},r\rightarrow\infty)=0.
\]
This continuous Schwarzschild time at the boundary serves as the CFT
time and will be denoted by $t$ or $t'$ in what follows. \\

What is the dual boundary picture that corresponds to this Vaidya
geometry? The dual boundary process is that of a thermalization following
a ``sudden quench''. One turns on a source in the CFT that couples
to a relevant operator, for a very short duration. The source pumps
in some finite amount of energy, say $E$, into the CFT and sends
it to an excited level, the energy density of the excited boundary state is $O(N)$ since it is dual to a large black hole in the bulk. Thereafter the excited but isolated CFT ``relaxes''
i.e. the energy $E$ gets ``equipartitioned''\footnote{Here in the shell collapse example, the ``equipartition" is not happening in position space as the perturbation is spherically symmetric in the bulk and consequently energy injection is uniform all over the boundary. But this spreading happens in energy space, involving levels around $E$.} through strong nonlinear (chaotic)
self-interactions of the CFT \cite{feingold1986distribution} in the limit of large system size i.e.
$N\rightarrow\infty$. This is the so called Pure state or Eigenstate
thermalization phenonmenon \cite{PhysRevA.43.2046,PhysRevE.50.888,rigol2008thermalization}. They are
hypothesized under very broad and generic conditions, to explain the
observation that systems which are initially prepared in far off-equilibrium
states, tend to evolve in time to a state which appears to be in thermal
equilibrium as a result of chaotic/ non-linear interactions. This is
currently a highly active topic of investigation
and its implications for AdS/CFT and gravity is being explored as
well \cite{Khlebnikov:2013yia}. We give a concise summary of this in the appendix \ref{sec:A-Pr=0000E9cis-on ETH}.\\

%%%%%%%%%%%%%%%%%%%%%%%%%%%%%%%%%%%%%%%%%%%%%%%%%%%%%%%
\subsection{Bulk reconstruction in the Vaidya-SAdS geometry\protect \\
}
%%%%%%%%%%%%%%%%%%%%%%%%%%%%%%%%%%%%%%%%%%%%%%%%%%%%%%%

Now we construct boundary smearing representation of normalizable
local bulk (supergravity) operators for the Vaidya-AdS spacetime. In the
Vaidya case, although $v$ is global coordinate, the metric is non-analytic
in $v$. The global vacuum i.e. $v$-vacuum in the collapse geometry,
unlike in the case for the Kruskal (Hartle-Hawking) vacuum in the
eternal SAdS spacetime, is not preserved under any isometry.\\

For points to the future of the shell, i.e. $v>0$ or regions $I$
and $II$, we can pretend that we are in an eternal SAdS$_{d+1}$
geometry. In particular, for region $I$, local bulk operator insertions
can be reconstructed from the CFT data by using the smearing functions
for the corresponding region of the eternal AdS given by (\ref{eq: Smearing for SAdS_d+1 outside horizon})
(obtained by \cite{Kabat:2014kfa}) with support on the spacelike separated
region of the (complexified) conformal boundary. This is demonstrated
in figure ~\ref{fig:Vaidya-AdS}, the yellow segment of the boundary supports the smearing
for encoding a local bulk insertion at the point $P$ in region $I$.
The smearing expression is identical to (\ref{eq: Smearing for SAdS_d+1 outside horizon})
but we rewrite it here in a form with new labels for convenience,
\begin{equation}
\Phi(P)=\int_{\mbox{spacelike}}dt'd\theta'd\Omega'_{d-2}\; K_{SAdS}(P|t',\theta',\Omega'_{d-2})\;\mathcal{O}(t',\theta',\Omega'_{d-2}).\label{eq: SAdS smearing function}
\end{equation}
%Evidently the subscript $I$ denotes fields inserted in region $I$.
Similarly for a local bulk operator inserted at a point $S$ in region
$III$ in the diagram, due to the fact that the entire spacelike separated
region with respect to $S$ is pure AdS, it is very easy to write down
a smearing expression using the pure AdS smearing function \cite{Hamilton:2006az,Hamilton:2006fh}:
\[
\Phi(S)=\int_{\mbox{spacelike}}dt'd\theta'd\Omega'_{d-2}\; K_{AdS}(S|t',\theta',\Omega'_{d-2})\:\mathcal{O}(t',\theta',\Omega'_{d-2}).
\]
Recall that $\theta$ is imaginary. \\

However, in general for operator insertions in regions $II$, $III$
and $IV$, one cannot simply apply this recipe. In general cases the
lightcones bounding spacelike regions may cross the shell and one
needs to continue/ propagate modes across the shell to reach the right
boundary (see figure ~\ref{fig:Vaidya-AdSR}). Such a construction was
carried out for the AdS$_{2}$ shell collapse in \cite{Lowe:2008ra}
but that construction was greatly aided and simplified by the fact
that in $1+1$ spacetime dimensions, the geometry is always locally
pure AdS$_{2}$ and hence one can, in essence, use the AdS$_{2}$
spacelike Green's functions to construct the smearing function in
all parts of the spacetime%
\footnote{Such a result, obviously would apply for $2+1$ dimensions as well,
since even in $2+1$ dimensions, the metric is also always locally
pure AdS$_{3}$. In fact the AdS$_{2}$ construction can be understood
as a dimensional reduction of the AdS$_{3}$ collapse (reduction of
three dimensional general relativity to two-dimensional Jackiw-Teitelboim
gravity).%
}. Below we will consider each region in increasing order of complexity in their
boundary smearing reconstruction.\\

%%%%%%%%%%%%%%%%%%%%%%%%%%%%%%%%%%%%%%%%%
\subsection{Points to the past of the shell (region $IV$)\protect \\
}
%%%%%%%%%%%%%%%%%%%%%%%%%%%%%%%%%%%%%%%%%

\begin{figure}[htbp] 
\begin{center} 
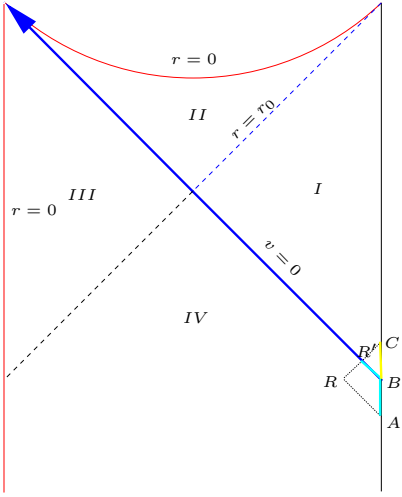 
\caption{Smearing regions for operator insertions to the past of shock.} 
\label{fig:Vaidya-AdSR} 
\end{center}
\end{figure}

For region $IV$, such as a point $R$ in figure~\ref{fig:Vaidya-AdSR}
where a local bulk operator is inserted, we can again use a spacelike
Green's function for pure AdS, say $G_{AdS}$ \cite{Hamilton:2006az,Hamilton:2006fh}
to express the bulk operator as a sum of an integral over the boundary
(segment $AB$) and an integral over the surface of the shock on the
$v\rightarrow0^{-}$ (segment $BR'$) using Green's theorem \cite{Lowe:2008ra};
\begin{eqnarray*}
\Phi_{R} & = & \int_{v'=0^{-}}\sqrt{h|_{v=0^{-}}}dr'd\Omega'_{d-1}\:\left[\Phi(v',r')\nabla_{v}G_{AdS}(R|v',r')-G_{AdS}(R|v',r')\nabla_{v}\Phi(v',r')\right]\\
 &  & \qquad+\int_{r'\rightarrow\infty}\sqrt{h|_{r'\rightarrow\infty}}dv'd\Omega'_{d-1}\:\left[\Phi(v',r')\nabla_{r'}G_{AdS}(R|v',r')-G_{AdS}.(R|v',r')\nabla_{r}\Phi(v',r')\right].
\end{eqnarray*}
This is indicated in cyan on figure~\ref{fig:Vaidya-AdSR}. Of course
the second integral, i.e. the one over the asymptotic boundary is
converted to a smearing function integral (as was done in \cite{Lowe:2008ra})
using the fall-off behaviors of $\Phi$ and $G_{AdS}$ as the $\sqrt{h}G\partial_{r'}\Phi$
term vanishes in the limit $r'\rightarrow\infty$. However one needs
some work on the integral expression over the surface of the shock,
$BR'$ to turn it into an integral over the asymptotic boundary, $BC$.
The first step in this direction is to use appropriate matching conditions
to translate fields and their (normal) derivatives from the $v\rightarrow0^{-}$
side to the $v\rightarrow0^{+}$ side. The matching conditions for
the field and the normal derivatives of the field can be easily obtained
by integrating the Klein-Gordon equation across the shock. Since there
are no sources on the shock, both the field and its normal derivative
are continuous across the shock.
\begin{eqnarray}
\Phi(v\rightarrow0^{-},r,\Omega) & = & \Phi(v\rightarrow0^{+},r,\Omega),\nonumber \\
\nabla_{v}\Phi|_{v\rightarrow0^{-}} & = & \nabla_{v}\Phi|_{v\rightarrow0^{+}}.\label{eq: continuity across the shock}
\end{eqnarray}
Thus we have now, the intermediate expression,
\begin{eqnarray*}
\Phi_{R}(v,r,\Omega_{d-1}) & = & \int_{v'=0^{+}}\sqrt{h}dr'd\Omega'_{d-1}\:\left[\Phi(v',r')\nabla_{v}G_{AdS}(R|v',r')-G_{AdS}.(R|v',r')\nabla_{v}\Phi(v',r')\right]\\
 &  & \qquad+\int_{v}^{0}dv'\int d\Omega'_{d-1}\: K_{AdS}(t_{i}(v),r,\Omega_{d-1}|t'_{i}(v'),\Omega'_{d-1})\mathcal{O}_{\Delta}(t'{}_{i}(v'),\Omega'_{d-1}).
\end{eqnarray*}
Then the second step is to use the smearing representation for fields
outside horizon in a SAdS background i.e. (\ref{eq: Smearing for SAdS_d+1 outside horizon}),
for the fields on the $v\rightarrow0^{+}$ side of the shock. So we
arrive at the following expression for the field as a smeared operator,
\begin{eqnarray}
\Phi_{R} & = & \int_{0}^{t(C)}dt''d\Omega''_{d-1}K'_{SAdS}(R|t'',\Omega''_{d-1})\:\mathcal{O}_{\Delta}(t'',\Omega''_{d-1})\nonumber \\
 &  & \qquad+\int_{v}^{0}dv'\int d\Omega'_{d-1}\: K_{AdS}(t_{i}(v),r,\Omega_{d-1}|t'_{i}(v'),\Omega'_{d-1})\mathcal{O}_{\Delta}(t'{}_{i}(v'),\Omega'_{d-1}).\label{eq: Smearing for points in region III like R}
\end{eqnarray}
where
\[
K'_{SAdS}(R|t'',\Omega''_{d-1})=\int_{v'=0^{+}}\sqrt{h}dr'd\Omega'_{d-1}\:\left[K_{SAdS}(t_{f}(v'),r',\Omega'_{d-1}|t'',\Omega''_{d-1})\overleftrightarrow{\nabla}_{v}G_{AdS}(R|v',r')\right].
\]
Here $t(C)$ is the time coordinate for the point $C$ on the boundary in figure~\ref{fig:Vaidya-AdSR}.\\

%%%%%%%%%%%%%%%%%%%%%%%%%%%%%%%%%%%%%%%%%%%%%%%%%%%%%%%%%%%%%%%%%%%
\subsection{Points inside the black hole horizon (region $II$)\label{sub:Points-inside-the horizon}\protect \\
}
%%%%%%%%%%%%%%%%%%%%%%%%%%%%%%%%%%%%%%%%%%%%%%%%%%%%%%%%%%%%%%%%%%%

\begin{figure}[htbp] 
\begin{center}
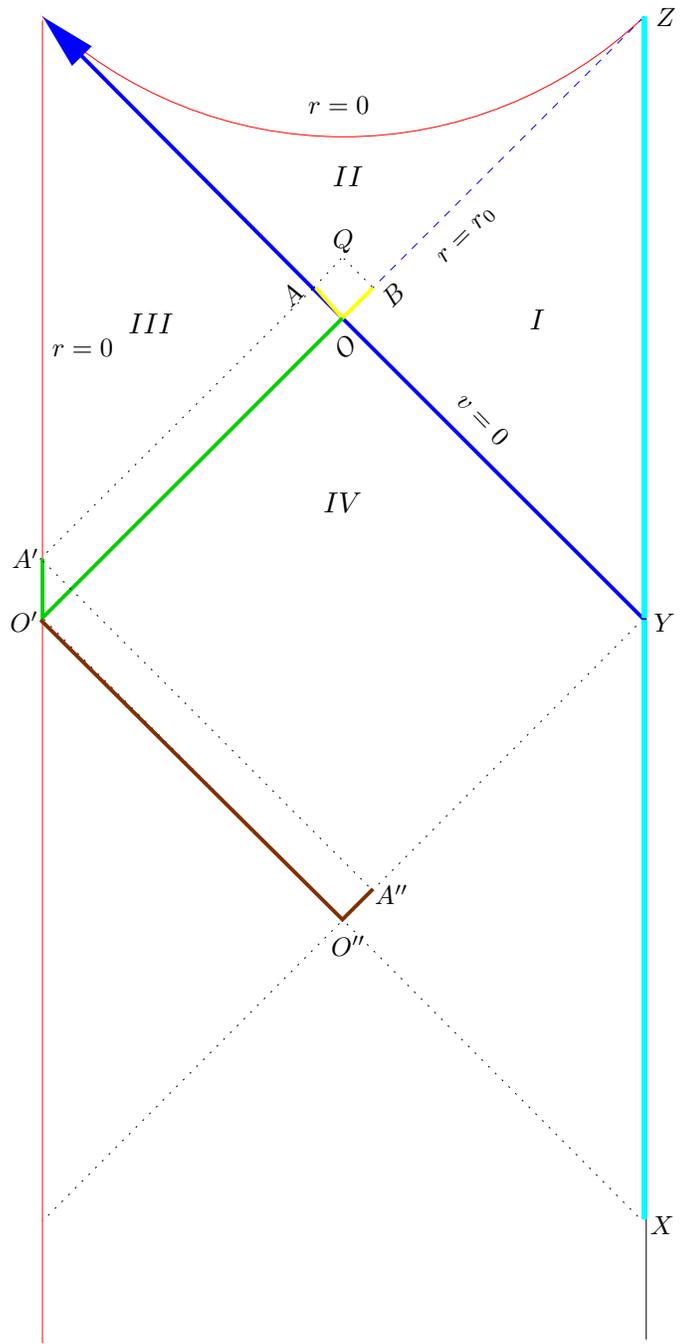 
\caption{The Penrose diagram depicting smearing domains for points inside the black hole horizon, such as $Q$.} 
\label{fig:Vaidya-AdSQ}
\end{center}
\end{figure}

For local bulk operator insertions in region $II$ such as $Q$ in
figure ~\ref{fig:Vaidya-AdSQ}, one needs to proceed in the following
steps.\footnote{For a similar back-evolving construction for BHs at infinite $N$, see \cite{Heemskerk:2012mn}.}\\

\begin{enumerate}
\item Use a \emph{retarded} Green's function in a Schwarzschild AdS geometry,
i.e. $G_{SAdS}$, to express a field at $Q$ as sum over an integral
over the future horizon and an integral over the surface of the shock
on the side of region $II$ i.e. $v'\rightarrow0^{+}$

\footnote{In doing this we are directly encountering the so called trans-Planckian problem \cite{Almheiri:2013hfa,Papadodimas:2015jra, Harlow:2014yka}, the energy of collision of the back-moving scalar particle from $Q$ with the shell at $v\rightarrow0^{+}$ spills over into energies higher than any UV cut-off scale. However recall that a) at this point we are at strictly infinite $N$ or infinite Planck energy, and even at finite $N$ we are assuming that b) the infalling shell is not charged under the scalar field, so collision does not transfer energy. We revisit this issue in section \ref{sec:Outlook} where we demonstrate that, despite the fact that our argument here fails when the field interacts with gravity, we can provide an adhoc working formula in section \ref{sec:Outlook} even in the presence of gravitational interactions. We do not have a more satisfactory resolution at this moment.}.

 This is indicated
by the segments $AO$ and $OB$ in yellow in figure~\ref{fig:Vaidya-AdSQ}.
\begin{eqnarray*}
\Phi_{Q} & = & \int_{v'=0^{+}}\sqrt{h}dr'd\Omega'_{d-1}\:\left[\Phi(v',r')\nabla_{\perp}G_{SAdS}(Q|v',r')-G_{SAdS}.(Q|v',r')\nabla_{\perp}\Phi(v',r')\right]\\
 &  & +\int_{r'=r_{0}}\sqrt{h}dv'd\Omega'_{d-1}\:\left[\Phi(v',r')\nabla_{\perp}G_{SAdS}(Q|v',r')-G_{SAdS}.(Q|v',r')\nabla_{\perp}\Phi(v',r')\right].
\end{eqnarray*}

We have introduced $\nabla_{\perp}$ as a shorthand for ``normal derivative" and we shall stick to this notation henceforth to reduce clutter.

\item Use matching conditions, (\ref{eq: continuity across the shock})
to express the field and its derivative in the integral
over the shock on the $v\rightarrow0^{+}$ side to those on the $v\rightarrow0^{-}$
side i.e. the locally pure AdS side. 
\begin{eqnarray*}
\Phi_{Q} & = & \int_{v'=0^{-}}\sqrt{h}dr'd\Omega'_{d-1}\:\left[\Phi(v',r')\nabla_{\perp}G_{SAdS}(Q|v',r')-G_{SAdS}.(Q|v',r')\nabla_{\perp}\Phi(v',r')\right]\\
 &  & +\int_{r'=r_{0}}\sqrt{h}dv'd\Omega'_{d-1}\:\left[\Phi(v',r')\nabla_{\perp}G_{SAdS}(Q|v',r')-G_{SAdS}.(Q|v',r')\nabla_{\perp}\Phi(v',r')\right].
\end{eqnarray*}

\item For the fields in the integral over the future horizon, i.e. segment
$OB$, now use the SAdS smearing function, equation (\ref{eq: Smearing for SAdS_d+1 outside horizon})
to express it as an integral of the timelike boundary segment $YZ$
in figure~\ref{fig:Vaidya-AdSQ}.
\begin{eqnarray}
\Phi_{Q} & = & \int_{v'=0^{-}}\sqrt{h}dr'd\Omega'_{d-1}\:\left[\Phi(v',r')\overleftrightarrow{\nabla}\,_{\perp}G_{SAdS}(Q|v',r')\right]
 +\int_{0}^{\infty}dt_{f}d\Omega''_{d-1}\:\mathcal{O}_{\Delta}(v'',\Omega''_{d-1})\nonumber \\
 &  & \left(\int_{r'=r_{0}}\sqrt{h}dv'd\Omega'_{d-1}\:\left[K_{SAdS}(v',r',\Omega'_{d-1}|v'',\Omega''_{d-1})\overleftrightarrow{\nabla}_{\perp}G_{SAdS}(Q|v',r',\Omega'_{d-1})\right]\right).\nonumber\\\label{eq: Phi_Q semi-complete}
\end{eqnarray}

\item Now, the integral over the surface of the shock, $AO$ has to be manipulated
into an integral over the timelike boundary in the right. Again, to
accomplish this we use the retarded AdS Green's function to map it
on the surfaces indicated by the green segments $A'O'$ ( i.e $r=0$)
and $OO'$. So we have, 
\begin{eqnarray*}
& &\Phi(v\rightarrow0^{-},r',\Omega')\nonumber\\
 & = & \int_{r'=0}\sqrt{h}dv''d\Omega''_{d-1}\:\Phi(v'',r''=0,\Omega''_{d-1})\overleftrightarrow{\nabla}\,_{t}G_{AdS}(v'\rightarrow0^{-},r',\Omega'_{d-1}|v'',r''=0,\Omega''_{d-1})\\
 &  & +\int_{r=r_{0}}\sqrt{h}dv''d\Omega''_{d-1}\:\left[\Phi(v'',r''=0,\Omega''_{d-1})\overleftrightarrow{\nabla}\,_{\perp}G_{SAdS}(v'\rightarrow0^{-},r',\Omega'_{d-1}|v'',r'',\Omega''_{d-1})\right].
\end{eqnarray*}

\item The integral over the Green segment, $OO'$ which is a null segment
on pure AdS, can be expressed as a sum of integrals over the boundary
portion, $XY$ and the surface of the shock on the pure AdS side i.e.
$v\rightarrow0^{-}$, if we substitute the fields appearing on the
segment $OO'$ by data on $XY$ and $OY$ using a spacelike
AdS Green's function, $G_{AdS}$ \cite{Hamilton:2005ju,Hamilton:2006az,Hamilton:2006fh}.
The $XY$ integral is of course a smearing integral. 
\begin{eqnarray*}
\Phi(OO') & = & \int_{XY}\: K_{AdS}(OO'|XY)\;\mathcal{O}_{\Delta}(XY)\\
 &  & +\int_{OY}\:\Phi(OY)\overleftrightarrow{\nabla}\,_{\perp}G_{AdS}(OO'|OY).
\end{eqnarray*}
Here to reduce clutter we have adopted a condensed notation where,
the coordinates of a point on a segment, say $OO'$ is suppressed
and represented also by $OO'$, i.e. $(v,t,\Omega){}_{OO'}\equiv OO'.$
\item The second integral in step 5, over $OY$ can be expressed as a smearing
integral over $YZ$ in the following way. First we use continuity
of fields across the shock to turn the integral into an integral on
the SAdS side i.e. $v\rightarrow0^{+}$. Then we replace those fields
by their SAdS smearing representation on the boundary domain, $YZ$.
\[
\int_{OY}\:\Phi(OY)\overleftrightarrow{\nabla}\,_{\perp}G_{AdS}(OO'|OY)=\int_{YZ}\:\mathcal{O}(YZ)\:\left(\int_{OY}\: K_{SAdS}(OY|YZ)\:\overleftrightarrow{\nabla}\,_{\perp}G_{AdS}(OO'|OY)\right).
\]
Thus combining the results of steps 5 and 6 we have reduced the integral
over $OO'$ in step 4 to a boundary integral over the boundary domain,
$XZ$.
\item Next, we map the $A'O'$ integral to a sum of integrals over the brown
segments $O'O''$ and $A''O''$ by using the retarded AdS Green's
function to replace the fields on $A'O'$ to data on $O'O''$ and
$A''O''$ 
\begin{eqnarray*}
\Phi(v'',r''=0,\Omega'') & = & \int_{O'O''}\,\Phi(O'O'')\overleftrightarrow{\nabla}\,_{\perp}G_{AdS}(v'',r''=0,\Omega''|O'O'')\\
 &  & \qquad+\int_{A''O''}\:\left[\Phi(A''O'')\overleftrightarrow{\nabla}\,_{\perp}G_{SAdS}(v'',r''=0,\Omega''|A''O'')\right].
\end{eqnarray*}

\item  Finally, we use the pure AdS smearing function to rewrite the integral
over the segment $O''A''$ to a boundary integral over the domain,
$XY$. And, analogous to the way we did for $OO'$, we now rewrite
the integral over the $O'O''$ as a boundary integral over the segment
$XZ$.
\[
\Phi(O'O'')=\int_{t(X)}^{\infty}dt\int d\Omega\: K_{AdS}(O'O''|t,\Omega)\: \mathcal{O}_{\Delta}(t,\Omega),
\]
\[
\Phi(A''O'')=\int_{t(X)}^{0}dt\int d\Omega\: K_{AdS}(A''O''|t,\Omega)\: \mathcal{O}_{\Delta}(t,\Omega).
\]
Here again we have used a condensed notation.
\item Finally we plug the expressions of steps 8 into step 7 and which in
turn is plugged into the step 4 and that in turn is plugged in equation
(\ref{eq: Phi_Q semi-complete}) . Since all these nested substitutions
make the final appearance of $\Phi_{Q}$ very cumbersome, we just
represent here schematically the result,
\begin{equation}
\Phi(Q)=\int_{v_{min}}^{0}dt\, d\Omega\: K^{-}(Q|t,\Omega)\:\mathcal{O}(t,\Omega)+\int_{v}^{\infty}dt\: d\Omega\: K^{+}(Q|t,\Omega)\:\mathcal{O}(t,\Omega).\label{eq: Subhorizon final schematic}
\end{equation}
where $v_{min}=t(X)$, time coordinate of the point $X$ on the boundary. $K^{\pm}$ denotes the smearing functions supported to the future/ past of the shell.
\end{enumerate}

%%%%%%%%%%%%%%%%%%%%%%%%%%%%%%%%%%%%%%%%%%%%%%%%%%%%%%%%%%%%%%%
\subsection{Points to the past of the shell with $r<r_{0}$ (region $III$)\protect \\
}
%%%%%%%%%%%%%%%%%%%%%%%%%%%%%%%%%%%%%%%%%%%%%%%%%%%%%%%%%%%%%%%%

\begin{figure}[htbp] 
\begin{center} 
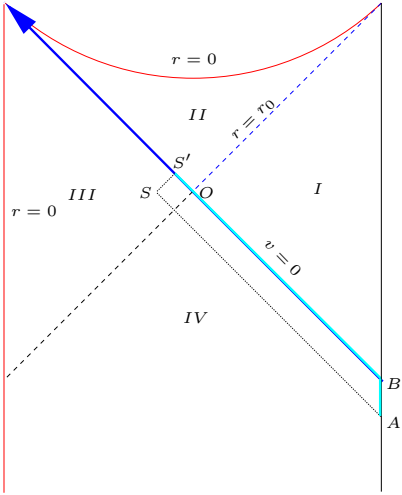 
\caption{Smearing regions for bulk operator insertions in the trapped region to the past of shock.} 
\label{fig:Vaidya-AdSS} 
\end{center}
\end{figure}

Such a bulk operator insertion is indicated at the point $S$ in the
figure ~\ref{fig:Vaidya-AdSS}. We will be completely schematic about
this case as all relevant mathematical expressions can be reduced
to those which have already been covered in the previous cases. At
the first step, we proceed identically as we did for an insertion at the point
$R$ in region $IV$ in figure ~\ref{fig:Vaidya-AdSR}, i.e. express
the field as an integreal over the shock $BS'$ and a part of the
boundary to the past of $v=0$, i.e. $AB$ using Green's theorem. Then
the integral over the shock, i.e. $BS'$ needs to be split up into
an integral of the parts of the shock inside and outside the horizon,
i.e. $OS'$ and $BO$ respectively. Then for $BO$ and $AB$ we can
proceed again identically as we did for the boundary construction
of the data on $BR'$ and $AB$ in figure ~\ref{fig:Vaidya-AdSR}, while
for the $OS'$ integral we can proceed as in step 4 of subsection \ref{sub:Points-inside-the horizon}
for the segment $OA$ in figure ~\ref{fig:Vaidya-AdSQ}.\\

%%%%%%%%%%%%%%%%%%%%%%%%%%%%%%%%%%%%%%%%%%%%%%%%%%%%%%
\section{Finite $N$ effects: early and late time cut-offs on the smearing
functions}\label{sec:VaidyaAdS_finteN}
%%%%%%%%%%%%%%%%%%%%%%%%%%%%%%%%%%%%%%%%%%%%%%%%%%%%%%

\begin{figure}[htbp] 
\begin{center} 
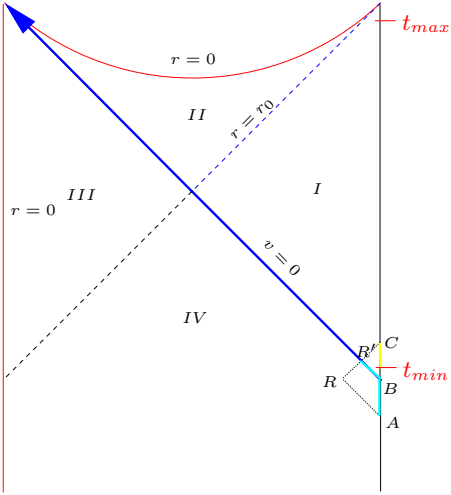 
\caption{Smearing regions for operator insertions to the past of shock at finite $N$. The smearing integral domain to the past of the shock in cyan i.e. $AB$, remains unchanged, while the integral over $BC$ is cut off at the lower limit i.e. $t_{min}$ (yellow segment). In the event when $R$ approaches the horizon, $C$ will shift higher up in the boundary and we shall have to truncate the integral at $t_{max}$. } 
\label{fig:Vaidya-AdSRC} 
\end{center}
\end{figure}

In this section we consider the effects of the finiteness of $N$
on the question of reconstructing local bulk observables from the
boundary data. Here our strategy is to follow the general principle
outlined in \cite{Kabat:2014kfa}. They found that for finite but
large $N$, it is indeed possible to reconstruct approximately local
bulk and the deviation from exact locality is non-perturbatively small
in $N$, i.e. $\sim e^{-N}$. This slight departure from locality
holds for operator insertions near to or within the black hole horizon.
Their approximately local construction was based on a very specific
course graining procedure which is designed to overlook late time
effects in a finite $N$ CFT. Late time effects in a finite $N$ CFT
can overhaul thermal behavior. In the smearing construction this translates
to imposing a late time cut-off, $t_{max}$ in the smearing integral,
\[
t_{max}\propto\frac{N}{\Delta}.
\]
Beyond this time the CFT correlators, so to say, ``recoheres'' away
from its ``thermal'' behavior and it ruins bulk locality completely.
This coarse-graining procedure consists of removing such parts of
the CFT correlator unlike the more common form of coarse graining
where we usually ``average over'' microstates. In case of the eternal SAdS black hole it suffices
to remove the late time parts of the CFT, but in our case of a pure
state black hole formed by a quench, this late time cut-off is not
enough. One also needs to throw away the ``early time'' part of
the CFT correlators when they are yet to loose the memory of their
initial pure state. This initial time, say $t_{min}$, prior to which the pure state or
eigenstate does not appear to be well approximated by a mixed thermal state is not very well
understood or well reported quantity in the literature. Since ETH
is still a relatively new field, only few studies have appeared which
have considered the effects of finite $N$ 
\cite{ikeda2013finite,beugeling2014finite,beugeling2014off,shchadilova2014quantum,alba2014eigenstate,khodja2015relevance,danshita2014quantum,kim2014testing}.
However one can try to make an estimate of this time.
Since this time is a decoherence time because observables lose phase
information associated with off-diagonal terms {(}see e.g. appendix \ref{sec:A-Pr=0000E9cis-on ETH}{)},
it is expected to be exponentially suppressed with size of the Hilbert space or number density of states around the energy $E$, $n(E)=e^{S(E)}$ \footnote{Note that in fermionic systems, the number of states/levels around energy $E$ increases with the total number of atoms/degrees of freedom due to level-splitting as a result of Pauli exclusion principle. So while keeping the subsystem size fixed, increasing the size or number of degrees of freedom of complement results in increase of density of states and hence improves decoherence. The complement acts as a better and better heat reservoir.}. Near
thermal equilibrium, the dissipation timescales are set by dimensional
reasons in a CFT to be that of inverse temperature. So we arrive at
a heuristic expression for the decoherence time to be,\footnote{Note that $t_{min}$ is not a time scale at which point the correlation function behaves badly or shows a behavior of thermality. Rather this time scale is important from the point of view of smearing function, which we assume to be vanishing before this time scale. In some sense, this can be understood as quantifying our ignorance about the precise microscopic structure constituting the horizon and beyond (which could be something like fuzzball e.g.) and rather estimating the non-localities one has to introduce to obtain an approximate semiclassical description behind horizon even at finite $N$. This is in line with the cut-off applied in \cite{Kabat:2014kfa}, where again, it was not the correlation function alone which is badly behaved after this time scale, but rather the whole convolution of the correlation function with the smearing function.}
\begin{equation}
t_{min}\sim\frac{e^{-\alpha S}}{\beta}.\label{eq: thermalization time initial time cut off}
\end{equation}
Here $\alpha$ is some positive number. Now, in the infinite $N$
limit, this time evidently vanishes and the CFT thermalizes instantly.
However at finite but large values of $N$ this time is small but
still non-vanishing. Prior to this time, CFT correlators show strong
depedence on the initial pure state as well, as they are yet to ``decohere''
i.e. yet to lose crucial phase information for off-diagonal terms.
Bulk reconstruction expressions from the last section must now be
modified. All integrals from time coordinates $\left(0,\infty\right)$,$(0,v)$
and $(v,\infty)$ must be replaced by the ranges $\left(t_{min},t_{max}\right)$,
$(t_{min},v)$ and $(v,t_{max})$ respectively. For example, bulk
operators in region $II$ such as an operator inserted at $R$, the
infinite $N$ expression (\ref{eq: Smearing for points in region III like R})
must now be modified: 
\begin{eqnarray*}
\Phi_{R}^{N} & = & \int_{t_{min}}^{t_{max}}dt''d\Omega''_{d-1}K'_{SAdS}(R|t'',\Omega''_{d-1})\:\mathcal{O}_{\Delta}(t'',\Omega''_{d-1})\\
 &  & \qquad+\int_{v}^{0}dv'\int d\Omega'_{d-1}\: K_{AdS}(t_{i}(v),r,\Omega_{d-1}|t'_{i}(v'),\Omega'_{d-1})\mathcal{O}_{\Delta}(t'{}_{i}(v'),\Omega'_{d-1}).
\end{eqnarray*}
We indicate this modified smearing in figure~\ref{fig:Vaidya-AdSRC}. Here we note two points:
\begin{enumerate}
\item We did not insert a cut off in the integration domain which is to
the past of the quench or shock i.e. $v\rightarrow0^{-}$ in the second
term, but we did so in the first term, $v\rightarrow0^{+}$, because
the CFT decoheres \emph{after} the quench.
\item For insertion points, $R$ which are close to the boundary, the domain
of integration never reaches $t_{max}$, and in that case the upper
cut-off is redundant. However when $R$ approaches the horizon, then
the smearing domain might extend beyond $t_{max}$ and hence needs
to be cut off.
\end{enumerate}

\begin{figure}[htbp] 
\begin{center}
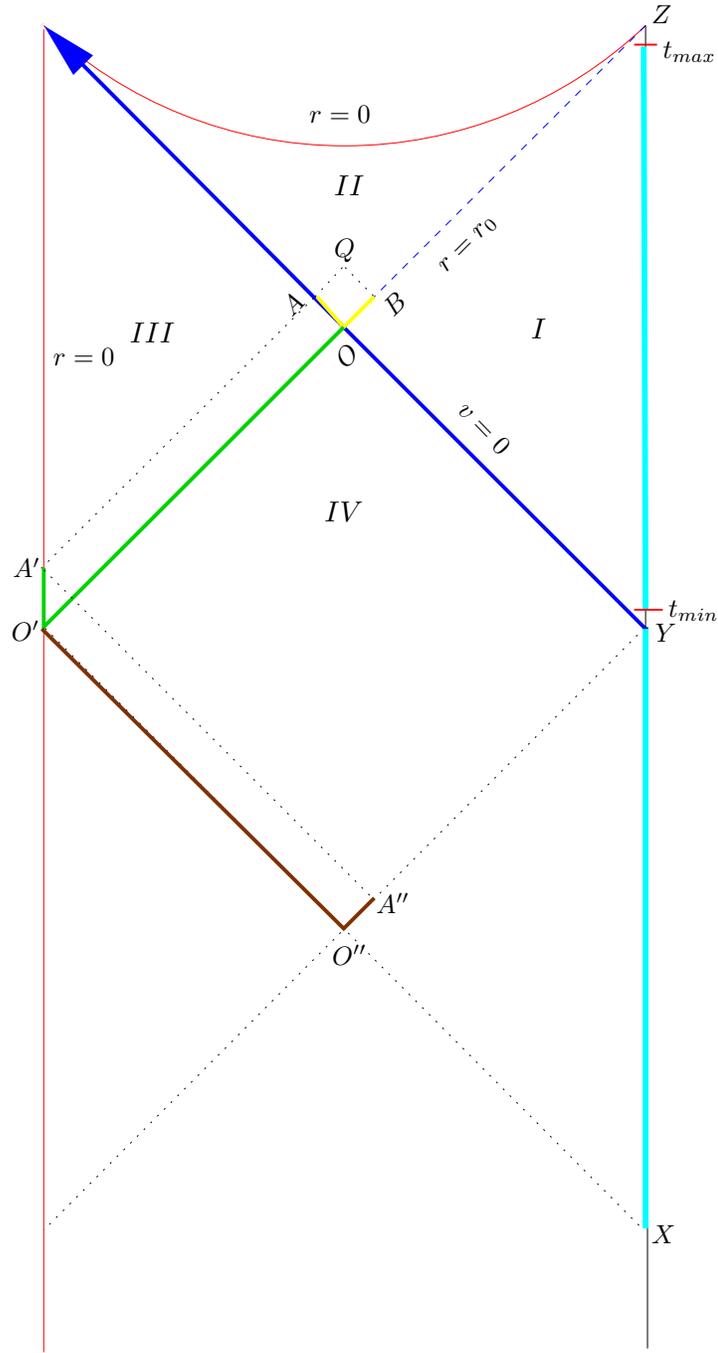 
\caption{Finite $N$ modifications to the smearing domains for points inside the black hole horizon, such as $Q$. The part of smearing integral indicated by the cyan segment (which previously covered the entire $YZ$ at infinite $N$) is now cut off at both early and late times.} 
\label{fig:Vaidya-AdSQC}
\end{center}
\end{figure}

We are especially interested in operator insertions inside
the horizon such as at points like $Q$ of figure~\ref{fig:Vaidya-AdSQ}
at finite $N$. For this case, we need to impose both early and late
time cut-offs, since the smearing domain extends across the shock,
i.e. $t=0$ and all the way to future infinity. This is displayed
in figure~\ref{fig:Vaidya-AdSQC}. Analogous modifications would need to be done for insertions in the
trapped region, $III$.\\

It is interesting to estimate the departure from bulk locality as
a result of this early time cut-off. Since we are excluding the region
from $t=0$ to $t=t_{min}$ which is a very small region, we can approximate
this excluded contribution by the size of the region, $\Delta t=(t_{min}-0)$
times the value of the integrand $I$ at the lower limit i.e. $t=0$. 
\[
\int_{0}^{t_{min}}dt\, d\Omega\: I(t,\Omega)\approx t_{min}\:\int d\Omega\: I(0,\Omega)\;\dots\propto\frac{e^{-\alpha S}}{\beta}.
\]
This will be true for any integrals, including that of commtuators
of spacelike separated bulk operators. So the change in locality i.e. in the
commutators would be $\sim e^{-\alpha N}$ which is non-perturbatively
small in $N$ (in the black hole case, $S\propto N$).

%%%%%%%%%%%%%%%%%%%%%%%%%%%%%%%%%%%%%%%%%%%%%%%%%%%%%%%
\section{Discussions and relation to other approaches\protect \\}\label{sec:discuss}
%%%%%%%%%%%%%%%%%%%%%%%%%%%%%%%%%%%%%%%%%%%%%%%%%%%%%%%

In this concluding section, we want to explore the implications of our bulk reconstruction by smearing functions vis-à-vis the black hole information paradox. This paradox is one of the long standing and most studied problems in recent times. Specifically after the discovery of the \emph{firewall paradox} \cite{Almheiri:2012rt},\cite{Braunstein:2009my}, the original paradox of Hawking has been further sharpened. Various proposals in favor and against this firewall argument have been suggested and the existing literature on this is already quite vast which we do not list out here. In this section, we spell out the status of our framework by comparing and contrasting it with few existing proposals of bulk reconstruction for black holes in AdS.\\

The firewall proposal originally arose from trying to simultaneously satisfy the following semiclassical notions in black hole physics and finds a negative answer\footnote{More precisely, it arose from trying to reconcile strong subadditivity property of entanglement entropy between three quantum systems during a black hole evaporation process: early Hawking radiation (radiation emitted from a black hole before Page time), late radiation residing near but outside the horizon and Hawking modes just inside the horizon.}
\begin{enumerate}
\item Unitarity of the physical process (in other words, existence of a unitary $S$-matrix) during black hole formation and its eventual evaporation,
\item Validity of semiclassical physics in the near and outside horizon region of the black hole,
\item To an external observer, the black hole is a quantum system with discrete spectrum and
\item Validity of equivalence principle near the horizon which, in other words, is the existence of the smooth horizon to an infalling observer.
\end{enumerate}

While firewall goes for the breakdown of the last scenario, an alternative way out is the violation of semiclassical physics in the near horizon region. Although the local curvature near a macroscopic black hole is low enough to neglect quantum gravity corrections, it should be noted that the violation of semiclassical physics can arise from precisely such small but \emph{nonlocal} corrections going as $1/N$ or $e^{-N}$. In fact most of the recent resolutions hinging on this breakdown of semiclassical physics is manifested by the dependence of early radiation on the modes just inside the horizon. Such non-locality could be arising from a smooth wormhole type structure between these two modes \cite{Maldacena:2013xja} or just due to the non-perturbative effects \cite{Hamilton:2007wj},\cite{Kabat:2014kfa},\cite{Papadodimas:2012aq},\cite{Papadodimas:2013jku}. Such non-locality was also a key aspect of the previous black hole complementarity picture \cite{Susskind:1993if}. \\

Below, we will mostly concern ourselves with this above mentioned non-locality induced by $1/N$ or $e^{-N} $ effects. For example, Papadodimas-Raju (PR) \cite{Papadodimas:2012aq},\cite{Papadodimas:2013jku} and even earlier works \cite{Hamilton:2007wj},\cite{Kabat:2014kfa} explicitly constructed such non-local operators in the framework of AdS/ CFT\footnote{Thus in these references and in the present work we primarily address AdS black holes, while we anticipate similar resolutions for more phenomenological situations in flat spacetime. An alternate (and somewhat complementary to the non-local business) resolution to firewall would be that at the late times, the quantum gravity path integral is dominated by geometries without a horizon (the modification of horizon due to such $1/N$ or $e^{-N}$ factors as mentioned above)  \cite{Solodukhin:2005qy},\cite{Germani:2015tda}.}, which implies that, when recast in terms of early modes and inside modes, these two physical systems are actually not independent.

%%%%%%%%%%%%%%%%%%%%%%%%%%%%%%%%%%%%%%%%%%%%%%%%%%%%%%%%%%%%%%%%%%
\subsection{Comparison with Papadodimas-Raju proposal}
%%%%%%%%%%%%%%%%%%%%%%%%%%%%%%%%%%%%%%%%%%%%%%%%%%%%%%%%%%%%%%%%%%

So far we have discussed the most natural way (as it seems to us) to extend the smearing construction to finite $N$, for local bulk operators behind the horizon of pure state black holes. As mentioned before, an alternative construction has been done by PR, where such sub-horizon bulk operators were constructed for both one-sided and two-sided black holes. We find it particularly useful to discuss their approach in some detail as it sheds light on how to apply these non-locality techniques to the firewall problem. This prescription involves the necessity for introducing a second set of CFT operators $\tilde{\mathcal{O}}$, besides the usual quasi-primary operators $\mathcal{O}$ that we discussed before. These $\tilde{\mathcal{O}}$ operators were called ``mirror operators" which are defined by certain properties or rules when they act on a particular CFT state. In PR construction, these $\mathcal{O}$ and $\tilde{\mathcal{O}}$ operators and their relation to the bulk fields are expressed in the momentum space, where they behave like the usual creation-annihilation operators. At large $N$ it is clear that for two-sided eternal black hole geometries, for both the smearing and PR prescriptions to be equivalent, the set of operators on the left side of the CFT should be related (but not same) to the requisite mirror operators \cite{Papadodimas:2012aq}. It is a priori not clear whether any such connections are present for the present case, i.e. AdS-Vaidya geometry. Our analytically continued ``left" CFT and the left operators can only be defined in near thermal states (states with energy density $O(N)$), not for any arbitrary CFT state. Incidentally, the PR mirror operators can be explicitly constructed for the geon geometries \cite{Guica:2014dfa} mentioned before and presumably also similarly for eternal SAdS geometries.\\

For one sided black holes at finite $N$, these two approaches have quite a few commonalities and differences. First of all, both the constructions have a strong notion of ``background-dependence'' built into them. This is already incorporated in PR as well, which also has a notion of ``state dependence''.\footnote{The status of state dependence is currently under investigation, especially whether it might imply an ``observable'' violation of Born rule in quantum mechanics. We won't be elaborating on this issue further. For more details see e.g. \cite{Papadodimas:2015jra}.} The issue of state dependence is very well discussed in PR, where the mirror operators satisfy some required properties on a given CFT state $|\Psi\rangle$. In position space, the background dependence can be understood as the presence of different smearing function construction for different background spacetimes \cite{Papadodimas:2013jku}. It is not quite clear whether HKLL prescription is necessarily state dependent and so far the answer seems to be negative.\\

State dependent property of both this smearing function and PR construction plays a very important role in circumventing the no-go theorems against the possibility of such sub-horizon operator reconstruction from the CFT \cite{Almheiri:2012rt},\cite{Almheiri:2013hfa},\cite{Marolf:2013dba}. They have been nicely summarized in \cite{Papadodimas:2013jku}. However, there are further arguments against the state dependent construction for non-equilibrium states as discussed in e.g. \cite{Bousso:2013ifa} and \cite{Harlow:2014yoa}. It remains an important question on how to fully incorporate a non-equilibrium phenomena in such state dependent construction (although for a recent discussion along this line see \cite{Papadodimas:2015jra}). It also remains to be seen whether it is at all possible or meaningful to construct a state independent `smearing operator' as argued in one of our earlier paper \cite{Kabat:2012hp}.\\

However, the real difference between the smearing approach and the PR mirror construction is in how to treat the finite $N$ non-local effects mentioned before. One needs to remember that the firewall proposal and their resolutions deal with a situation where the quantum gravity effects are manifest due to the finiteness of the Hilbert space (fields have finite matrix dimensions $N$) or entropy of the CFT $S$. However, $N$ is also large enough so that one can still talk about small quantum gravity effects around a semiclassical backgrounds without going to a full, unknown non-local quantum ``geometry''. In PR construction, the finite $N$ effects come into play when we go beyond the lower point functions in the bulk or the boundary, where the number of operator insertions $n$ is much smaller with respect to the CFT central charge. If $n$ is large enough (for an estimate see \cite{Papadodimas:2013jku}, \cite{Harlow:2014yoa}), then the `edge effects' lead to the breakdown of large $N$ factorization property of the CFT state and hence of a local bulk description. This type of non-local effects are visible also in the smearing construction and they are the (non-perturbative in) $1/N$ corrections to the thermal state. As discussed in \cite{Kabat:2011rz}, even though it is possible to construct a local bulk operator at infinite $N$ by adding a tower of smeared higher dimensional multi-trace boundary operators, this procedure fails for finite $N$ due to stringy exclusion principle \cite{Maldacena:1998bw}.
%\footnote{It must be noted that the parametric dependence on $N$ in this case may not just depend on the $1/N$ factors appearing outside the next to leading order terms in the large $N$ factorization. In particular, it is also important to understand how does the number of operators at every order, scale with $N$ at finite $N$. This might change the overall $N$ dependence of these terms. Also, this particular type of corrections are present even in empty AdS case. Here in particular, we have considered BHs which saturates the holographic entropy bound.}
But in our approach the non-perturbative effects are included by making our bulk insensitive to the precise microstates or precise information about the collapsing pure state. Namely the position space smearing representation of the bulk operator explicitly shows the importance of the late time behavior of the CFT correlation function towards the construction of locality. This requires one to put the late time cut-off on the correlation functions, which in turn makes the bulk fields non-perturbatively non-local (by an order $e^{-S/2}$) \cite{Kabat:2014kfa}. The fact that these late time cut-offs are extremely important for studying the physics close to the horizon and hence for issues regarding firewalls, is not clearly or directly visible in the momentum space and other similar constructions where the support of the bulk to boundary transfer function is extended over the entire boundary regardless of the location of the resulting bulk operator close to or far away from the horizon. \\

However, as mentioned in \cite{Kabat:2014kfa}, this might suggest the appearance of a firewall after `Page time' for a large non-evaporating AdS black hole \footnote{Although large black holes do not evaporate, they do radiate Hawking quanta and reabsorb the emitted quanta. It is fair to call $t_{max}\sim \beta S$ as the Page time because by this time, the black hole has radiated (and reabsorbed) half its entropy and in principle an observer measuring these quanta can extract one single bit of information. It can be checked that this general expression, $\beta S$, in the small AdS black hole limit reduces to the more familiar expression of the Page time scale of order $M^3$ we associate with asymptotically flat black hole (in 3+1-dimensions).

Bulk insertions in the black hole exterior which are to the future of $t_{max}$ cannot be reconstructed as their entire CFT smearing domain (spacelike separated boundary points) fall in the cut-off/excision region. Moreover, it is interesting to note that this is precisely the time scale where the many saddle point sum approach in the case of AdS$_3$ breaks down \cite{Kleban:2004rx}.} when an infalling observer can in principle measure an order $e^{S/2}$ numbers of commutators. \cite{Kabat:2014kfa} show that circa $t_{max}$, correlators, and hence commutators, in a finite $N$ CFT begin to deviate from thermal behavior by an amount of order $e^{-S/2}$ and cumulative deviation due to the ability to measure $e^{S/2}$ commutators is $O(1)$. This $O(1)$ quantum correction at the horizon can be viewed as a firewall. However, this appearance of the firewall could just be an artifact of only considering the semiclassical saddle point geometries \cite{Germani:2015tda}. To truly avoid firewall, it might be essential to consider geometries without a horizon which might get important at later times in the quantum gravity path integral. \\

Note that even though we simply extend the infinite $N$ smearing construction for sub-horizon operator insertions to finite $N$, without independently arguing against the objections raised in e.g. \cite{Almheiri:2013hfa} or \cite{Marolf:2013dba}, our construction is completely fine. The reason here is that the background dependent nature of the construction of the smearing function precisely avoids these objections, just like PR construction, and simultaneously shows the effects of both $1/N$ and $e^{-S}$ towards bulk locality. Whereas PR extends the local operator construction for pure state cases by arguing for the presence of Mirror operators in any CFT, here we achieve the construction by incorporating either a `second set' of operators by analytically continuing them in imaginary time or by using a combination of retarded AdS Green's function and the smearing function.  It remains to be seen, whether this procedure could be equivalent to the construction of Mirror operators in any sense. \\

%%%%%%%%%%%%%%%%%%%%%%%%%%%%%%%%%%%%%%%%%%%%%%%%%%%%%%%
\section{Outlook\protect \\}\label{sec:Outlook}
%%%%%%%%%%%%%%%%%%%%%%%%%%%%%%%%%%%%%%%%%%%%%%%%%%%%%%%

There are couple of issues that we would want to understand better in the future. As we mentioned before, some of the most immediate problems are to understand the state dependence and their role in information problem. It would also be pleasing to see how to understand or incorporate the effect of stringy exclusion principle in violating locality. Such effects are also non-perturbative in $1/N$ expansion, but their precise formulation still needs to be clarified.\footnote{There are non-local effects which are perturbative in $1/N$ due to interactions but we do not need to worry about them as the non-locality they induce can be removed by redefining the bulk operator, i.e. by adding smeared multi-trace operators order by order in $1/N$ to the leading planar single trace operator \cite{Kabat:2011rz}.} A natural question is how or if the stringy exclusion principle lead to the same effects as to putting a late (and/ or early) time cut-off in the CFT. One other immediate issue is the trans-Planckian problem. There are two related issues here. First issue is regarding the collision of the back-evolving field quanta with the infalling shell at trans-Planckian energies, briefly alluded to in footnote \ref{transplanckian} \cite{Almheiri:2013hfa,Harlow:2014yka}, which comes back if the bulk fields we considered are the same fields which constitute the matter of the shell falling in to create the black hole in the first place. The second issue is that to back-evolve sub-horizon bulk fields and map them to region $I$, leads to the precarious step of using trans-Planckian energies due to the red-shifting effect of the horizon. Restricting to low (sub-Planckian) energies, as it should be for field theory in curved spacetime to be valid, means limiting the bulk reconstruction of sub-horizon operators to a timescale after formation of the black hole to Schwarzschild
times of order $\beta$log$ S$, where $S$ is
the Bekenstein-Hawking entropy \cite{Papadodimas:2015jra, Avery:2015hia}. This is of course the black hole ``scrambling time" \cite{Sekino:2008he,Susskind:2011ap}. Then our Green's function approach for bulk reconstruction certainly encounters a trans-Planckian problem for post-scrambling sub-horizon insertions. At infinite $N$ of course $S$ is infinite and we can describe the full interior, but at finite $N$ reconstruction of interiors of horizon is thwarted beyond the scrambling time. However, we briefly note here that this is not an insurmountable difficulty. If instead of using the retarded Green's function method of \ref{sub:Points-inside-the horizon}, we use the smearing function for sub-horizon insertions for post-scrambling times like $Q$ using ``complexified time" expression (\ref{eq. No trans-Planckian}), we never encounter trans-Planckian energies either in the collision or in region $I$. Thus we arrive at a very nice picture for the collapse black hole. Before the passage of one scrambling time after the formation of the horizon/black hole, the collapse black hole interior is different from the eternal black hole as evident in the Green's functions used to back-evolve fields into region $I$. But after the passage of time of the order of the scrambling time after the formation of the black hole, the interior is approximated better by the interior of eternal black hole. This is of course a superficial argument to sidestep the trans-Planckian issue, perhaps deserving more scrutiny, and we would like to think of this as still an open issue.

Finally, there is the more interesting case of an evaporating or small black hole. There we are dealing with a situation with a metastable state of false vacuum in the CFT, which ultimately decays to a stable state (thermal AdS). Our work in this paper does not seem to address that situation as we have not taken into account the backreaction from the emitted Hawking quanta. However we believe that a similar progress can be made in developing CFT construction/description by proceeding along the lines of \cite{Lowe:2013zxa, Lowe:2014vfa}.

%%%%%%%%%%%%%%%%%%%%%%
\section*{Acknowledgments}
%%%%%%%%%%%%%%%%%%%%%%

We thank Dan Kabat for numerous helpful discussions and for giving his valuable feedback on the draft. SR thanks David Lowe for perceptive comments and explaining features of his highly relevant works \cite{Lowe:2013zxa, Lowe:2014vfa, Avery:2015hia}. SR also thanks Martin Kruczenski for patiently explaining in detail several points in his interesting work on ETH \cite{Khlebnikov:2014gza,Khlebnikov:2013yia}. DS thanks Gia Dvali, Monica Guica and Michael Haack for discussions. The work of SR is partially supported by the American-Israeli Bi-National Science Foundation, the Israel Science Foundation Center of Excellence and the I Core Program of the Planning and Budgeting Committee and The Israel Science Foundation ``The Quantum Universe". DS is supported by the ERC Self-completion grant. SR also thanks Shibaji Roy on behalf of the Theory Division, SINP, Kolkata for his warm hospitality for a period during which this work was initiated.

\section*{APPENDIX}

\appendix

\section{A Précis on Eigenstate thermalization\label{sec:A-Pr=0000E9cis-on ETH}}

The idea of eigenstate thermalization was introduced to explain relaxation
in isolated quantum systems following a large perturbation or sudden
injection of large amount of energy. The idea is that in an isolated many body system
put in a pure state, (coarse-grained) observables such as correlation
functions of operators which are supported on a (small) proper subset
of the system ``thermalizes'' to their microcanonical ensemble correlation
function\footnote{For a CFT$_d$ in volume $V$, we have $S \sim \left(c V E^{d-1} \right)^{\frac{1}{d}}$, where $c$ is some generalized notion of central charge in $d$-dimensions which also serves as the planar factorization parameter, $c \sim N^2$. This is a generalization of the Cardy formula for the asymptotic density of states of a 1+1-dimensional CFT  to arbitrary spacetime dimensions. This implies the microcanonical ensemble at energy $E$ is equivalent to the canonical ensemble at inverse temperature $\beta \sim \left( cV/E \right)^{\frac{1}{d}}.$ This can be derived from the relationship of the Hawking temperature and the mass of the SAdS black hole with its dual CFT on a $S^{d-1}$ (with radius $R$, i.e. the bulk AdS radius) and at a temperature equal to the Hawking temperature of the black hole \cite{Witten:1998zw}.}. A heuristic way of motivating this phenomenon is by considering
a subsystem or a subset of degrees of freedom and likening the complement
of the subsystem to act as a ``heat bath'' with respect to the subsystem.
The interactions of the subsystem with the complement turn the reduced
density matrix of the subsystem to appear arbitrarily close to the
canonical or suitable grand canonical ensemble to an arbitrary degree
of accuracy in inverse powers of the total number of degrees of freedom,
say $N$. So in the large $N$ limit, any subsystem with ``size''
parametrically smaller than $N$ would appear same as their thermal
averages \cite{2013arXiv1307.0378L}. The whole system of course is still in a pure state and initial state information is hidden in cross-correlations between observables/operators with support on the subsystem and observables with support in the complement. In particular this pure state can even be an energy eigenstate
of the full $N$-body Hamiltonian, hence the name ``eigenstate''
thermalization. If the system is integrable then the subsystems are
in some sense free or independent and the system never equilibrates
or relaxes. Consider a system of $N$ degrees of freedom placed in
a pure state, not necessarily an energy eigenstate, which is expressed
as follows in energy basis,
\[
|\Psi(0)\rangle=\sum_{i}|E_{i}\rangle\langle E_{i}|\Psi(0)\rangle.
\]
Here we are assuming the range of energies over which the state has
significant support is sufficiently narrow compared to the mean energy,
\[
\Delta E=\sqrt{\langle H^{2}\rangle-\langle H\rangle^{2}}\ll\langle H\rangle.
\]
This constrains the coefficients, 
\[
\frac{\sum_{i}E_{i}^{2}||\langle E_{i}|\Psi(0)\rangle||^{2}}{\left(\sum_{i}E_{i}||\langle E_{i}|\Psi(0)\rangle||^{2}\right)^{2}}-1\ll1.
\]
After time $t$ the state evolves to,
\[
|\Psi(t)\rangle=\sum_{i}|E_{i}\rangle\langle E_{i}|\Psi(0)\rangle e^{-iE_{i}t}
\]
Now the expectation value of a local operator at time $t$ from such
an initial state is,
\begin{eqnarray}
\langle\Psi(t)|\mathcal{O}|\Psi(t)\rangle & = & \sum_{i,j}\langle E_{j}|\mathcal{O}|E_{i}\rangle\langle E_{i}|\Psi(0)\rangle\langle E_{j}|\Psi(0)\rangle^{*}e^{-i\left(E_{i}-E_{j}\right)t}\nonumber \\
 & = & \sum_{i}\langle\mathcal{O}\rangle_{i}||\langle E_{i}|\Psi(0)\rangle||^{2}+\sum_{i\neq j}\langle E_{j}|\mathcal{O}|E_{i}\rangle\langle E_{i}|\Psi(0)\rangle\langle E_{j}|\Psi(0)\rangle^{*}e^{-i\left(E_{i}-E_{j}\right)t}.\nonumber\\\label{eq: ETH precursor}
\end{eqnarray}
Now at this point both the diagonal term as well as the off-diagonal term
look to be completely dependent on initial pure state, indicated by
the explicit dependence on the expansion coefficients $\langle E_{i}|\Psi(0)\rangle$.
Only the second term is however time-dependent. If this term, being a sum of a large number of random phases, vanishes due to mutual cancellations, then we are left with
contributions from diagonal elements (this is known as the \emph{diagonal ensemble}
in the statistical physics literature):
\[
\langle\Psi(t)|\mathcal{O}|\Psi(t)\rangle\approx\sum_{i}\langle\mathcal{O}\rangle_{i}||\langle E_{i}|\Psi(0)\rangle||^{2},
\]
which is \emph{still} an explicit function of the initial state and
does not look at all like a canonical ensemble average. Here we need
the second ingredient or assumption of the eigenstate thermalization
hypothesis - the correlators/observable $\mathcal{O}$ \emph{must be}
smooth function of energy (expectation value):

\[
\langle\mathcal{O}\rangle\sim\langle\mathcal{O}(\langle E\rangle)\rangle
\]
Then, a small spread in energy of the original initial state implies
a small spread in $\mathcal{\langle\mathcal{O}\rangle}$ as well
\[
\langle\mathcal{O}\rangle_{i}\sim \langle\mathcal{O}(\langle E\rangle\rangle +O(E_{i}-E).
\]
Using this and omitting the time-dependent off diagonal terms in (\ref{eq: ETH precursor}),
we get,
\[
\langle\Psi(t)|\mathcal{O}|\Psi(t)\rangle\sim\mathcal{\langle\mathcal{O}\rangle}(\langle E\rangle)+O(E_{i}-E).
\]
Thus the pure state expectation value ``relaxes'' to the microcanonical
average as soon as the off-diagonal
time-dependent terms in (\ref{eq: ETH precursor}) become negligible.
We shall call this time duration (following a violent perturbation of the system),
the \emph{ETH thermalization} or \emph{ETH decoherence} time, $t_{min}$.
This nomenclature is justified since after this time-scale the (sub)system
loses crucial phase information contained in the off-diagonal terms,
equivalent to losing memory of the initial state. Once this time is past, the system is in the diagonal ensemble and by the second assumption of ETH, the diagonal ensemble for operators which is a smooth function of energy is the canonical ensemble. This second term in the rhs of (\ref{eq: ETH precursor}) being an interference or ``noise'' term, is inversely proportional to number of states in the range $\left( \langle E\rangle, \langle E\rangle+ \Delta E\right) $. Consequently one expects, the larger the density of states around the energy $\langle E \rangle$ the faster a subsystem would thermalize,

\[
t_{min} \sim e^{-\alpha S(\langle E \rangle)}, \alpha>0.
\]

\bibliographystyle{brownphys}
\bibliography{BH_finite_N1}

\end{document}